\documentclass[aps,pra,superscriptaddress,amsmath,amssymb,preprintnumbers,twocolumn,showpacs]{revtex4}
\usepackage{amssymb}
\usepackage{color}
\usepackage{graphicx}
\usepackage{amsmath, epsfig}

\begin{document}
\def\bbm[#1]{\mbox{\boldmath$#1$}}
\date{\today}

\title{Loss of coherence and dressing in QED}


\author{B. Bellomo}
\author{G. Compagno}
\affiliation{Dipartimento di Scienze Fisiche ed Astronomiche
dell'Universit\`{a} di Palermo, Via Archirafi, 36, 90123 Palermo,
Italy.}
\author{F. Petruccione}

\affiliation{School of Pure and Applied Physics Howard College,
University of KwaZulu-Natal Durban, 4041 South Africa.}


\begin{abstract}
The dynamics of a free charged particle, initially described by a
coherent wave packet, interacting with an environment, i.e. the
electromagnetic field characterized by a temperature $T$, is
studied. Using the dipole approximation the exact expressions for
the evolution of the reduced density matrix both in momentum and
configuration space and the vacuum and the thermal contribution to
decoherence, are obtained. The time behaviour of the coherence
lengths in the two representations are given. Through the analysis
of the dynamic of the field structure associated to the particle
the vacuum contribution is shown to be linked to the birth of
correlations between the single momentum components of the
particle wave packet and the virtual photons of the dressing
cloud.
\end{abstract}


\maketitle


\section{Introduction}

Decoherence consists in the destruction of coherences present in
the initial state of a quantum system due to the interaction with
external degrees of freedom \cite{Libro decoerenza 2002}.
Decoherence is associated to the increase of entropy and the loss
of purity of the initial state of the system
\cite{Palma-Suominen-Ekert 1996}.

The environment may be regarded as monitoring certain properties
of the quantum system through the interaction with the system
itself \cite{Zurek 2003}. Not all initial quantum states are
equally fragile to this interaction: often there are relatively
robust states with respect to it, called "pointer states"
\cite{Zurek 1981}. Experimental evidence of this environment
induced decoherence has also been recently reported \cite{Brune
1996,Myatt 2000, Brezger 2002,Auffeves 2003,Hackermüller 2004}.

In the case of a particle, either free or in a potential, linearly
coupled to the environment modelled as a bath of harmonic
oscillators at temperature $T$, several studies of decoherence
processes have already been reported  \cite{Hakim-Ambegaokar 1985,
Barone-Caldeira 1991, Ford 1993, Durr-Spohn 2000, Mazzitelli 2003,
Eisert 2004}. In these studies both the Hamiltonian approach and
functional techniques have been used. It has been shown that,
starting with the particle and the bath described by a factorized
density matrix, it is possible to distinguish two characteristic
contributions to the decoherence: the first related to the thermal
properties of the bath and the second, independent of temperature,
to the zero point fluctuations of the oscillators of the bath
\cite{Breuer-Petruccione 2001}. Decoherence has been shown for
charged particles initially described by a wave function made of a
coherent superposition of two moving wave packets to be linked to
the emission of Bremsstrahlung \cite{Petruccione-Breuer libro
2002}.

Here we want to investigate the role played by radiation emission
and entanglement with field degrees of freedom, on the decoherence
induced on a free charged particle by its interaction with the
electromagnetic field at temperature $T$ that plays the role of
environment. In particular we shall study the decoherence among
the components of an initially Gaussian free wave packet
representing the particle by analyzing the evolution of the off
diagonal elements of the particle reduced density matrix. We shall
start, as is typically done \cite{Feynman-Vernon
1963,Caldeira-Leggett 1985,Privman-Mozyrsky 1998,Tolkunov-Privman
2004}, from decoupled initial conditions which correspond to the
absence of initial correlations between the system and the
environment. Using physical approximations, we'll reduce the
particle-field interaction to a simple analytically solvable
model.

We will focus mainly on two aspects. The first one is the analysis
of the dynamics in two different basis. The aim is of evidencing
clearly how the change of representation gives place to different
relative importance of various effects induced by the coupling
with the environment. The second one is the study of build up of
quantum correlations between the system and the environment. Using
the fact that our model Hamiltonian allows exact treatment it is
easy to show in detail the mechanism linked to the part of
decoherence independent of temperature. This will be done by
investigating the time dependence of the effects on the particle
due to vacuum fluctuations, such as the dressing, and through the
analysis of bath dynamics without the use of approximations as the
Markovian one.

The paper is organized as follows. In Sec. \ref{par:Modello} we
describe the approximations adopted that transform the Hamiltonian
into a linear form amenable to exact treatment. In Sec.
\ref{par:Matrice densita ridotta} the particle density matrix is
obtained both in momentum and real space. In Sec. \ref{par:vacuum
contribution interpretation} we analyze the dynamics of the field
structure, evidencing the relationship between the part of
decoherence induced by vacuum and dressing process. In Sec.
\ref{par:Summary and Conclusions} we summarize and discuss our
results. In Appendixes \ref{funzione di decoerenza calcolo},
\ref{appendice trasformata}, \ref{dinamica campo} we have
collected most of the calculations to make more readable the main
body of the text.

\section{Model} \label{par:Modello}

The system under investigation is a free spinless particle of mass
$m_0$ and charge $e$ moving at initial velocity $\bbm[v]_0$,
interacting with the electromagnetic field in thermal equilibrium
at temperature $T$. The particle is initially described by a
coherent wave packet, whose initial width is assumed to be small
with respect to the relevant wave lengths of the electromagnetic
field. The interaction between the system and its environment is
described by the non relativistic minimal coupling Hamiltonian
with an upper cut off frequency $\Omega$ corresponding to a wave
length such that the dipole approximation may be applied
\cite{Petruccione-Breuer libro 2002}.

The adoption of dipole approximation is standard in the treatment
of free particle decoherence \cite{Barone-Caldeira 1991, Ford
1993, Durr-Spohn 2000} but it limits the validity range to times
of the order of $\tau_0 = c/v_0\Omega$, where $c$ is the light
speed and $v_0$ is the initial velocity of the particle. This
limitation can be made less strong by using a "moving dipole"
approximation which consists in substituting the particle position
operator $\bbm[\hat{r}]$ by a parameter $\bbm[r]_t$ indicating the
average wave packet position at time $t$. In absence of
interaction this is given by $\bbm[r]_t=\bbm[r]_0+\bbm[v]_0 t$,
$\bbm[r]_0$ being the initial position of the particle. It is
possible to check the consistency of our choice by comparing
$\bbm[r]_t - \bbm[r]_0$ with the particle average displacement in
presence of the interaction, $\langle\bbm[\hat{q}]\rangle_t$,
given by Eq.\,(\ref{mediaposizione2}). In fact their difference is
smaller than the wave packet width for times less than the ones
where moving dipole approximation can be applied (see Sec.
\ref{par:Summary and Conclusions}). Our results are valid until a
time $\tau_d$ such that because of  the spreading the wave packet
width becomes of the order of the minimal wave length involved in
the treatment \cite{Ford 1993}. The contribution to the spreading
of the wave packet due to the interaction can be shown to be for
small value of $\alpha$ (see Eq.\,(\ref{spread totale}))
negligible with respect to the free evolution for small times.
Taking an initial wave packet of minimum indetermination, using
Eq.\,(\ref{evoluzionelibera}) for the free spreading and the
dipole approximation condition $ \Delta r \ll c /\Omega $, we get
$ \tau_d \approx \Omega^{-1} (m_0 c)/\Delta p $ with $ \tau_d \gg
\tau_0$.

The potential vector in the Coulomb gauge is given by
\cite{Sakurai 1977}
\begin{equation}\label{potenziale vettore}
  \bbm[\hat{A}] (\bbm[r])=\sum_{k,j}
\bbm[\varepsilon]_{k,j}\sqrt{\frac{2\pi\hslash c^2}{V \omega_k
}}\left(\hat{\mathrm{a}}^{\dag}_{k,j}\mathrm{e}^{-i
\bbm[k]\cdot\bbm[\hat{r}]}+\hat{\mathrm{a}}_{k,j}\mathrm{e}^{i
\bbm[k]\cdot \bbm[\hat{r}]} \right) \, ,
\end{equation}
where $\bbm[\varepsilon]_{k,j}$ are the polarization vectors
$(j=1,2)$ of the mode $\bbm[k]$ of frequency $\omega_k$, periodic
boundary conditions are taken on a volume $V$, $\hslash$ is the
reduced Planck constant and $\hat{\mathrm{a}}_{k,j}$ and
$\hat{\mathrm{a}}^{\dag}_{k,j}$ are the annihilation and creation
operators of the field modes satisfying the commutation rules
$[\hat{\mathrm{a}}_{k,j},\hat{\mathrm{a}}^{\dag}_{k',j'}]=\delta_{kk'}\delta_{jj'}$.
The non relativistic minimal coupling Hamiltonian in the "moving"
dipole approximation is
\begin{flalign}\label{hamiltoniana
completa}
  \hat{H}=& \frac{1}{2m_0 }\left[\sum_p \bbm[p] \, \hat{\sigma}_p - \frac{e \bbm[\hat{A}]
  (\bbm[r]_t)}{c} \right]^2 +\sum_{k,j}\hslash \omega_k
  \hat{\mathrm{a}}^{\dag}_{k,j}\hat{\mathrm{a}}_{k,j}
  \nonumber \\= & \sum_{p} \frac{p^2}{2m_0}\;\mathbf{\hat{\sigma}}_p+\sum_{k,j}\hslash \omega_k
  \hat{\mathrm{a}}^{\dag}_{k,j}\hat{\mathrm{a}}_{k,j}  +\frac{e^2}{2m_0c^2}\bbm[\hat{A}]^2(\bbm[r]_t) \nonumber \\ & -
  \frac{e}{m_0c}\sum_{p} \mathbf{\hat{\sigma}}_p \bbm[p]\cdot \bbm[\hat{A}]
  (\bbm[r]_t) \, ,
\end{flalign}
where we have used $\bbm[\hat{p}]=\sum_p \bbm[p]\,\hat{\sigma}_p$,
$\hat{\sigma}_p=|p \rangle \langle p |$ is the projection operator
on the momentum $\bbm[p]$ and the potential vector is calculated
in $\bbm[r]_t$.

In Eq.\,(\ref{hamiltoniana completa}) the term quadratic in
$\bbm[\hat{A}]$, which is physically linked to the average
vibrational kinetic energy due to vacuum fluctuations
\cite{Weisskopf 1939}, can be exactly eliminated by a canonical
transformation of the Bogoliubov Tiablikov form
\cite{Hakim-Ambegaokar 1985}. Here it will be simply neglected
because it can be shown as usual to be very small compared to the
linear term.

Thus, using Eqs.\,(\ref{potenziale vettore}) and
(\ref{hamiltoniana completa}), the Hamiltonian reduces to the form
\begin{multline}\label{hamiltoniana}
  \hat{H}=\sum_p \frac{p^2}{2m_0}\;\hat{\sigma}_p+\sum_{k,j}\hslash \omega_k
  \hat{\mathrm{a}}^{\dag}_{k,j}\hat{\mathrm{a}}_{k,j} \\   +
  \sum_{p,k,j} g_{k,j}^p \hat{\sigma}_p \left(
  \hat{\mathrm{a}}^{\dag}_{k,j}\mathrm{e}^{-i \bbm[k]\cdot \bbm[r]_t}+\hat{\mathrm{a}}_{k,j}\mathrm{e}^{i \bbm[k] \cdot \bbm[r]_t}
  \right)\,,
\end{multline}
with the coupling coefficients given by
\begin{equation}\label{coefficiente di accoppiamento}
  g_{k,j}^p=-\bbm[p]\cdot
\bbm[\varepsilon]_{k,j}\frac{e}{m_0}\sqrt{\frac{2\pi\hslash}{V
\omega_k }} \,.
\end{equation}
Here, in contrast to other phenomenological models
\cite{Caldeira-Leggett 1983}, the coupling coefficients and the
spectral field properties are assigned, which allows to analyze
the dependence of the decoherence development on physical
parameters such as the mass and the charge of the particle.

The Hamiltonian of Eq.\,(\ref{hamiltoniana}) describing the
interaction between the system (particle) and environment
(electromagnetic field) is now treated exactly.

\subsection{System evolution} \label{par:Evoluzione del sistema}
In the interaction picture, introducing the time ordering operator
$T_{\leftarrow}$, the unitary time evolution operator is
\begin{equation}\label{operatore evoluzione}
  \hat{U}(t)=\mathrm{T}_{\leftarrow}\exp \left [-\frac{i}{\hslash}\int_0^t \mathrm{d}s
  \hat{H}_I(s) \right] \, ,
\end{equation}
where, from Eq.\,(\ref{hamiltoniana}), the interaction Hamiltonian
at time $t$ is given by
\begin{multline}\label{hamiltoniana d'interazione}
  \hat{H}_I(t)=\sum_{p,k,j}\hat{\sigma}_p
  g_{k,j}^p\left(
  \hat{\mathrm{a}}^{\dag}_{k,j}\mathrm{e}^{i (\omega_k-\bbm[k]\cdot\bbm[v]_0 )t}
  \mathrm{e}^{-i \bbm[k]\cdot\bbm[r]_0} \right.\\\left. + \hat{\mathrm{a}}_{k,j}
  \mathrm{e}^{-i (\omega_k-\bbm[k]\cdot\bbm[v]_0 )t}\mathrm{e}^{i \bbm[k]\cdot\bbm[r]_0} \right)\,.
\end{multline}
\\
The commutator of the interaction Hamiltonian at two different
times is equal to
\begin{equation}\label{commutatore}
  [\hat{H}_I(s),\hat{H}_I(s')]\!=\!-2i \! \sum_{p,k,j} g_{k,j}^{p\,2} \hat{\sigma}_p \sin \left[(\omega_k-\bbm[k]\cdot\bbm[v]_0 )
  (s-s')\right]
\end{equation}
where we have used
$\hat{\sigma}_p\,\hat{\sigma}_{p'}=\hat{\sigma}_p\delta_{pp'}$.
Because the commutator (\ref{commutatore}) commutes with the
interaction Hamiltonian, it is possible to give an exact
expression for the evolution operator \cite{Palma-Suominen-Ekert
1996,Petruccione-Breuer libro 2002} using the
Cambell-Baker-Hausdorf formula:

\begin{flalign} \label{operatore evoluzione cbh}
  \hat{U}(t)= & \exp \left[ -\frac{1}{2\hslash^2}\int_0^t \mathrm{d}s \int_0^t
  \mathrm{d}s'[\hat{H}_I(s),\hat{H}_I(s')]\theta (s-s')\right] \nonumber \\ & \times
  \exp \left[-\frac{i}{\hslash}\int_0^t \mathrm{d}s
  \hat{H}_I(s)\right]  \nonumber \\ = & \exp \left[i \sum_p \xi(p,t)\hat{\sigma}_p\right]
  \nonumber \\
  & \times \exp \left[\sum_{p,k,j}\hat{\sigma}_p g_{k,j}^p\left(
  \hat{\mathrm{a}}^{\dag}_{k,j}\alpha_k-\hat{\mathrm{a}}_{k,j}\alpha_k^*\right)\right]\,
  ,
\end{flalign}
where
\begin{equation}\label{alfa}
  \alpha_k=\frac{1-\mathrm{e}^{i(\omega_k-\bbm[k]\cdot\bbm[v]_0 )t}}{\hslash (\omega_k-\bbm[k]\cdot\bbm[v]_0 )}
  \mathrm{e}^{-i \bbm[k]\cdot\bbm[r]_0}\,.
\end{equation}
The term $\xi(p,t)$, present in the above phase factor, is a
number depending on the momentum $\bbm[p]$ and on the time $t$, as
it is shown in Appendix \ref{funzione di decoerenza calcolo}.

\section{Reduced density matrix analysis}
\label{par:Matrice densita ridotta}
The analysis of the decoherence of an initial coherent wave packet
will be conducted by examining the behaviour of the reduced
density matrix elements.

As initial condition we take a state with no correlation between
the particle and the electromagnetic field. To this condition
corresponds a decoupled initial density matrix of the form
\begin{equation}\label{stato iniziale}
  \hat{\rho} (0)=\hat{\rho}_S (0)\otimes \hat{\rho}_F \,,
\end{equation}
where $\hat{\rho}_S (0)$ represents the initially coherent wave
packet, while the field is taken in a thermal state at temperature
$T$ described by $\hat{\rho}_F=\exp(-\beta \hat{H}_F)/Z_F$, with
$\beta=1/k_BT$, $k_B$ the Boltzmann constant, $\hat{H}_F$ is the
Hamiltonian of the field and $Z_F$ the field partition function.

In Eq.\,(\ref{hamiltoniana}) the projection operator
$\hat{\sigma}_p$ commutes with $\hat{H}$, thus the particle's
momentum is a constant of motion. This implies that momentum space
provides a robust basis that allows to investigate easily the
decoherence development. Successively we shall consider the
coordinate space to see how the loss of coherence shows up in real
space.

\subsection{Momentum space} \label{par:Matrice
densita nello spazio dei momenti}

In the momentum representation the initial particle density matrix
becomes $\hat{\rho}_S (0)=\sum_{p,p'}\rho_S^{p,p'}(0)|p(0)\rangle
\langle p'(0)|$. Its elements at time $t$ are given by
\begin{flalign}\label{elementi matrice ridotta}
  \rho_S^{p,p'}(t)&=\langle p(t)|\hat{\rho}_S (t)|p'(t)\rangle  \\ &=\langle
  p(t)|\mathrm{tr}_F \{\hat{U}(t) \hat{\rho}_S (0)\otimes \hat{\rho}_F
  \hat{U}^{-1}(t)\}|p'(t)\rangle \nonumber \,,
\end{flalign}
where $|p(t)\rangle$ is an eigenstate of the momentum operator at
time $t$.

Indicating with $|\phi(t)\rangle$ an arbitrary field state we
obtain
\begin{multline}\label{evoluzione stato}
  \hat{U}(t)(|p(t)\rangle \otimes |\phi(t)\rangle) =  |p(t)\rangle \exp \left[i \xi(p,t)
  \right] \\ \otimes \exp \sum_{k,j}g_{k,j}^p\left(
  \hat{\mathrm{a}}^{\dag}_{k,j}\alpha_k-\hat{\mathrm{a}}_{k,j}\alpha_k^*\right)|\phi(t)\rangle
  \, ,
\end{multline}
where use has been made of the fact that the application of the
operator $\exp \left[i \sum_p \xi(p,t)\hat{\sigma}_p\right]$ of
$\hat{U}(t)$ on the state $|p(t)\rangle \otimes |\phi(t)\rangle$
leads to the factor $\exp \left[i \xi(p,t) \right]$. This factor
doesn't depend on the environment state but only on the associated
momentum.

We have already seen that with the Hamiltonian
(\ref{hamiltoniana}) the particle momentum is a constant of
motion. The states $|p(t)\rangle$ are stationary with respect to
the interaction and different momenta can't be connected by the
time evolution operator. Then, in Eq.\,(\ref{elementi matrice
ridotta}), in the momentum representation form of  $\hat{\rho}_S
(0)$, only the term $\rho_S^{p,p'}(0)|p(0)\rangle \langle
p'(0)|=\rho_S^{p,p'}(0)|p(t)\rangle \langle p'(t)| $ $\exp [
-it(p^2-p'^2)/2m_0\hslash]$ contributes to the reduced density
matrix evolution. Thus, Eq.\,(\ref{elementi matrice ridotta}) can
be written as
\begin{flalign}\label{elementi matrice ridotta2}
  &\frac{\rho_S^{p,p'}(t)}{\rho_S^{p,p'}(0)}=
\exp \left
\{i\left[\xi(p,t)-\xi(p',t)-\frac{t(p^2-p'^2)}{2m_0\hslash}\right]\right\}\times
\nonumber
   \\ &  \mathrm{tr}_F \left\{\exp \sum_{k,j}(g_{k,j}^p-g_{k,j}^{p'})\left(
  \hat{\mathrm{a}}^{\dag}_{k,j}\alpha_k-\hat{\mathrm{a}}_{k,j}\alpha_k^*\right)
  \hat{\rho}_F \right\}\,,
\end{flalign}
where we have used the property of ciclity of the trace.

We can rewrite this last expression as
\begin{equation}\label{funzione di decoerenza}
  \rho_S^{p,p'}(t)=\rho_S^{p,p'}(0) \exp
  \left[-\Gamma^{p,p'}(t)+i\Phi^{p,p'}(t)\right] \,,
\end{equation}
where we have introduced the decoherence function, typically used
in literature \cite{Petruccione-Breuer libro 2002,Ford 1993}, as
\begin{equation}\label{funzione di decoerenza calcolata2}
  \Gamma^{p,p'}(t)=-\sum_{k,j}\ln \mathrm{tr}_F
  \left\{\exp \left(
  \hat{\mathrm{a}}^{\dag}_{k,j}\gamma_{k,j}^{p,p'}-\hat{\mathrm{a}}_{k,j}\gamma_{k,j}^{p,p'}*\right)
  \hat{\rho}_F \right\}
\end{equation}
with $\gamma_{k,j}^{p,p'}=(g_{k,j}^p-g_{k,j}^{p'})\alpha_k$, and
the function $\Phi^{p,p'}(t)$
\begin{equation}\label{fattore di fase complessivo}
  \Phi^{p,p'}(t)=\xi(p,t)-\xi(p',t)
-\frac{t(p^2-p'^2)}{2m_0\hslash}\, ,
\end{equation}
that includes the phase term $\xi(p,t)-\xi(p',t)$ and the free
evolution term.

The decoherence function describes in a direct way the appearance
of decoherence. In fact, the increase of $\Gamma^{p,p'}(t)$ for
$\bbm[p] \neq \bbm[p]\,'$ gives rise to a decrease of the off
diagonal elements of the reduced density matrix, that is it leads
to the destruction of coherences among the different momenta in
the initial wave packet. Moreover, the expression of
$\gamma_{k,j}^{p,p'}$ shows that at $\bbm[p]=\bbm[p]\,'$ the
decoherence function is zero and then that the populations are
constant in time. This may be expected because, as shown, the
dipole approximation leads to momentum conservation.

For our model the calculation of the explicit form of the
decoherence function $\Gamma^{p,p'}(t)$ and the phase factor
$\xi(p,t)-\xi(p',t)$ is reported in Appendix \ref{funzione di
decoerenza calcolo}.

Eq.\,(\ref{parteraledecoerenzacontinuofinale}) shows that the
decoherence function $\Gamma^{p,p'}(t)$ increases quadratically
with the vector difference of the momenta $\bbm[p]-\bbm[p]\,'$.
Therefore there is decoherence in the off diagonal elements also
within the same energy shell. Introducing the spectral density,
\begin{equation}\label{spectral density}
    J(\omega)= \frac{2\alpha
}{3\pi}\frac{(\bbm[p]-\bbm[p]\,')^2}{m_0^2c^2} \,  \omega \exp
\left(-\frac{\omega}{\Omega}\right)\, ,
\end{equation}
containing the frequency dependent part of
Eq.\,(\ref{parteraledecoerenzacontinuofinale}) deriving from the
coupling coefficients and the density of the modes at frequency
$\omega$, $\Gamma^{p,p'}(t)$ can be rewritten as
\begin{flalign}
  \Gamma^{p,p'}(t)=
  \int_0^\infty \mathrm{d} \omega J (\omega) \frac{(1-\cos \omega t)}{\omega^2} \coth
  \left(\frac{\hslash \omega}{2k_BT}\right) \, .
\end{flalign}
Below the cut off frequency $\Omega$, $J(\omega)$ depends linearly
on $\omega$, this is typical of an Ohmic spectral density which
gives rise to frequency-independent damping
\cite{Petruccione-Breuer libro 2002}. This damping gives rise to a
loss of coherence between different momentum eigenstates but not
to dissipation, which is absent because the interaction
Hamiltonian commutes with the momentum operator.

In Appendix \ref{funzione di decoerenza calcolo} it is shown that
it is possible to separate in the decoherence function
$\Gamma^{p,p'}(t)$ the effects of vacuum fluctuations,
$\Gamma^{p,p'}_{vac}(t)$, and of thermal contribution,
$\Gamma^{p,p'}_{th}(t)$, as
$\Gamma^{p,p'}(t)=\Gamma^{p,p'}_{vac}(t)+\Gamma^{p,p'}_{th}(t)$.
Extracting the dependence on the momenta we rewrite the
decoherence function as
\begin{eqnarray}\label{decoerenza complessiva}
  \Gamma^{p,p'}(t)&= &\Gamma
(t)(\bbm[p]-\bbm[p]\,')^2   \\  & =&
  \frac{2\alpha }{3\pi}\frac{\ln\left[\sqrt{1 + \Omega^2t^2} \, \;\frac{\sinh(t/\tau_F)}
  {t/\tau_F} \right]}{m_0^2c^2}
    (\bbm[p]-\bbm[p]\,')^2\nonumber \,,
\end{eqnarray}
with $\Gamma (t)$ the decoherence factor and $\alpha=e^2/\hslash
c$ a dimensionless coupling constant. For the two contributions we
write
\begin{flalign}\label{contributo di vuoto}
  \Gamma^{p,p'}_{vac}(t)&= \Gamma_{vac}(t) (\bbm[p]-\bbm[p]\,')^2= \frac{2\alpha}{3\pi}\frac{\ln\sqrt{1 +
  \Omega^2t^2}}{m_0^2c^2} (\bbm[p]-\bbm[p]\,')^2 ,
\end{flalign}
with $\Gamma_{vac} (t)$ the vacuum decoherence factor and
\begin{flalign}\label{contributo termico}
  \Gamma^{p,p'}_{th}(t)
  &= \Gamma_{th}(t)(\bbm[p]-\bbm[p]\,')^2= \frac{2\alpha }{3\pi}\frac{\ln \left[ \frac{\sinh(t/\tau_F)}{t/\tau_F}   \right]}{m_0^2c^2}
   (\bbm[p]-\bbm[p]\,')^2  ,
\end{flalign}
with $\Gamma_{th} (t)$ the thermal decoherence factor and
$\tau_F=\hslash/\pi k_B T$ a characteristic thermal time. The
expression for $\Gamma^{p,p'}_{th}(t)$ is obtained under the
condition $k_BT\ll \hslash\Omega$. If $\hslash\Omega \approx
10^{-2} m_{\mathrm{e}}c^2$, $m_{\mathrm{e}}$ indicating the mass
of an electron, the above condition is well verified at ordinary
conditions ($T\ll 10^7 K$).

Eq.\,(\ref{decoerenza complessiva}) shows that $\Gamma^{p,p'}(t)$
increases faster with time, the difference $\bbm[p]-\bbm[p]\,'$
and the coupling constant $\alpha$.

From Eq.\,(\ref{fattore di fase}) we obtain:
\begin{equation}\label{decoerenza fase}
  \xi(p,t)-\xi(p',t)=\frac{2\alpha}{3\pi}\frac{p^2- p'^2}{m_0^2 c^2}(\Omega t - \arctan \Omega t
  )\,,
\end{equation}
which depends only on the energy difference between the components
of momentum $\bbm[p]$ rather than on their vector difference.
Separating the dependence from momenta we introduce from
Eqs.\,(\ref{fattore di fase complessivo}) and (\ref{decoerenza
fase}) the global phase factor $\Phi (t)$ as
\begin{eqnarray} \label{Bcompleto}
    \Phi^{p,p'}(t)&=&\Phi (t)\left(p^2- p'^2\right) \\ &=& \left[\frac{2\alpha \,(\Omega t-
  \arctan \Omega t)}{3 \pi   m_0^2 c^2} -\frac{t}{2m_0\hslash}\right]\left(p^2- p'^2\right) \nonumber \,.
\end{eqnarray}
We observe that $\Phi (t)$ doesn't depend on the initial state of
the field and that in absence of interaction it represents the
phase free evolution, $\Phi (t)=-t/2m\hslash$.

Using Eq.\,(\ref{decoerenza complessiva}) for $\Gamma^{p,p'}(t)$
and Eq.\,(\ref{Bcompleto}) for $\Phi^{p,p'}(t)$, we can rewrite
the particle density matrix elements of Eq.\,(\ref{funzione di
decoerenza}) as
\begin{equation}\label{matricemomenti}
  \rho^{p,p'}_S\!(t) =\rho^{p,p'}_S\!(0)\exp \!\! \left[-\Gamma (t) \left(\bbm[p]-\bbm[p]\,'\right)^2 \! + i
  \Phi (t)  \left(p^2-p'^2\right)\right] \!.
\end{equation}

To discuss the time evolution of the reduced momentum density
matrix elements it is useful to use simplified  expression for
$\Gamma (t)$ and $\Phi (t)$ for different times easily obtainable
from Eqs.\,(\ref{decoerenza complessiva}) and (\ref{Bcompleto}):
\begin{flalign} \label{decoerenza complessiva2}
&\Gamma (t) \:\approx\:\left\{\:\begin{aligned} \!\! \Gamma_{vac}
(t)\approx \frac{2\alpha}{3\pi m_0^2 c^2}\frac{\Omega^2 t^2}{2},
\quad\quad  \:\:\, \: \:\:\:\: t\ll\Omega^{-1} \,,
  \\ \\ \Gamma_{vac}
(t)\approx
 \frac{2\alpha
  }{3\pi m_0^2c^2} \ln \Omega t, \quad
  \,\Omega^{-1} \ll  t \ll  \tau_F \,,
  \\ \\ \Gamma_{th}
(t)\approx
 \frac{2\alpha
  }{3\pi m_0^2c^2}
  \frac{t}{\tau_F},
  \quad\quad\quad
  \:\: \: \:\:\:\, \: \:\: \: \:\: t \gg  \tau_F \,,
\end{aligned}
\right.
\end{flalign}
and
\begin{flalign} \label{fattore di fase approssimazioni}
&\Phi(t) \:\approx\:\left\{\:\begin{aligned} \!\! \frac{2 \,
\alpha }{3 \pi m_0^2 c^2} \frac{\Omega^3 t^3}{3}
    -\frac{t}{2m_0\hslash},\, \:\: \:  t\ll\ \Omega^{-1} \,,
  \\ \\
\frac{2\alpha}{3 \pi   m_0^2 c^2} \Omega t
  -\frac{t}{2m_0\hslash}, \, \: \qquad t\gg \Omega^{-1}.
\end{aligned}
\right.
\end{flalign}

We observe that the form of the decoherence factor $\Gamma (t)$
leads to a time behaviour for the reduced density matrix elements
analogous to the one obtained for an ensemble of two level systems
linearly interacting with a bath of harmonic oscillators
\cite{Palma-Suominen-Ekert 1996}. In our case there is an explicit
expression of the coefficients in terms of the parameters of our
system.

Using in Eq.\,(\ref{matricemomenti}) the approximated expressions
of $\Gamma (t)$, in the three time zones of Eq.\,(\ref{decoerenza
complessiva2}), and the expansion $\mathrm{e}^{-x}\approx 1-x$ for
$x\ll 1$ we obtain
\begin{flalign}\label{matrice densità varie zone}
 \left| \frac{\rho^{p,p'}_S(t) }{\rho^{p,p'}_S(0)} \right| \approx \left\{\:\begin{aligned}
\left[1-\frac{2\alpha}{3\pi}\frac{(\bbm[p]-\bbm[p]\,')^2}{m_0^2c^2}\frac{\Omega^2
t^2}{2}\right], \,\,  \:\: \:\; t\ll\Omega^{-1} \,,
  \\ \\
(\Omega t)^{-\frac{2\alpha
}{3\pi}\frac{\left(\bbm[p]-\bbm[p]\,'\right)^2}{m_0^2c^2}}
\;,\,\;\; \,\;\quad
  \Omega^{-1}\ll t \ll \tau_F\,,
  \\ \\
\exp\left[-\frac{2\alpha
}{3\pi}\frac{(\bbm[p]-\bbm[p]\,')^2}{m_0^2c^2}\frac{t}{\tau_F}\right],
 \quad\:\:\:\: t\gg \tau_F \,.
\end{aligned}
\right.
\end{flalign}
Eq.\,(\ref{matrice densità varie zone}) shows that the off
diagonal elements of $\rho^{p,p'}_S(t)$ evolve from the initial
value for small times with a quadratic trend, for intermediate
time with an hyperbolic and at large times with an exponential one
with the rate $2\alpha (\bbm[p]-\bbm[p]\,')^2/3\pi m_0^2c^2$.

\subsubsection{Vacuum and thermal contribution: decoherence times} \label{par:Tempi di decoerenza}
It is possible to use the approximated  expression of
Eq.\,(\ref{decoerenza complessiva2}) for $\Gamma (t)$ to evidence
the time regions in which vacuum and thermal contribution
dominate. It comes out that the vacuum contribution prevails for $
t \ll \tau_F$ while the thermal contribution dominates for $t\gg
\tau_F$. The transition time, $\tau_p$, at which the two
contributions are equal can be found imposing $ \ln \Omega \tau_p
= \tau_p / \tau_F $. This time doesn't depend on $\zeta = 2\alpha
(\bbm[p]-\bbm[p]\,')^2/3\pi m_0^2c^2$. For example for $\Omega
\approx 10^{19} s^{-1}$ ($\hslash \Omega \approx m_\mathrm{e}
c^2/100$) and $T=1 K$ we have  $\tau_F \approx 2.34 \cdot 10^{-12}
s$ from which we find $\tau_p \approx 10^{-10} s$.

In Fig. 1 the behaviour in time of $\Gamma^{p,p'}_{vac}(t)$ and
$\Gamma^{p,p'}_{th}(t)$ is shown as a function of physical
parameters present in $\zeta$. It shows that if
$\Gamma^{p,p'}_{vac}(t_p)\geq 1$ then vacuum contributes
effectively to decoherence, otherwise only the thermal
contribution will be effective.
\begin{figure}[h]\label{figura vuoto e termico}
\begin{center}
\includegraphics[width=0.46\textwidth, height=0.28\textwidth]{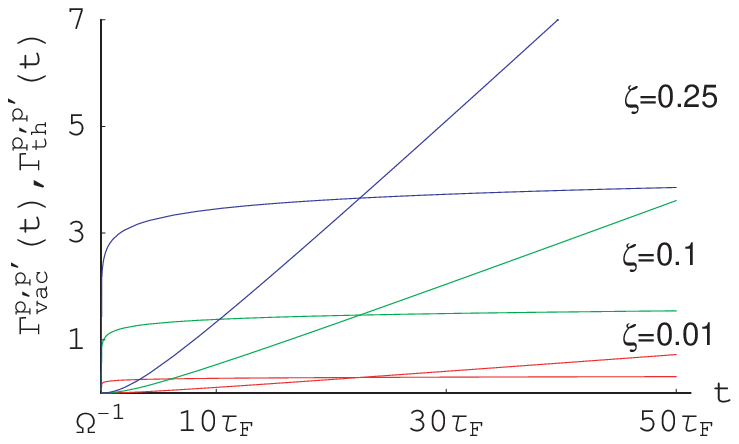}
\caption{\small{Figure shows $\Gamma^{p,p'}_{vac}(t)$ and
$\Gamma^{p,p'}_{th}(t)$ as a function of $\zeta = 2\alpha
(\bbm[p]-\bbm[p]\,')^2 / 3\pi
 m_0^2c^2$, in the case $T=300 K$. }}
\end{center}\end{figure}

In the range where the vacuum contribution dominates ($ t \ll
\tau_F$) there are two different typical time dependencies. In the
first one ($t\ll\Omega^{-1}$) the increase of decoherence is fast
while in the second one ($t\gg\Omega^{-1}$) it slows into a
logarithmic dependence. Fig. 2 represents the time development of
$\exp [-\Gamma^{p,p'}_{vac}(t)]$ as a function of the coupling
constant $\alpha$, showing that by increasing $\alpha$ and fixed
$\bbm[p]-\bbm[p]\,'$, we observe a decay of matrix elements due to
the vacuum contribution faster in time.
\begin{figure}[h]\label{figura2}  \centering
\includegraphics[width=0.45\textwidth, height=0.28\textwidth]{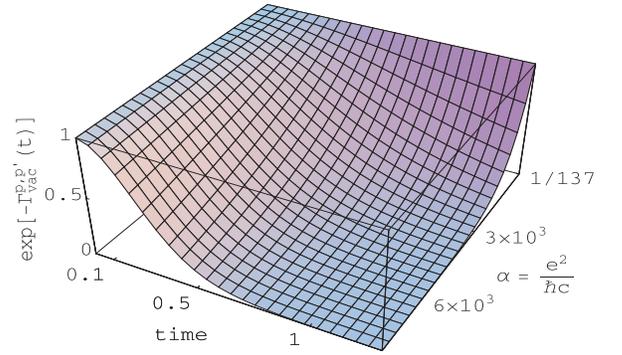}
\caption{\small{In figure it is plotted the time development of
$\exp [-\Gamma^{p,p'}_{vac}(t)]$ (\ref{contributo di vuoto}) as a
function of the coupling constant $\alpha=e^2/\hslash c$ in the
case $|\bbm[p]-\bbm[p]\,'|/m_0c = 0.1$. The time is taken in unit
of $t \Omega$ and the range of $\alpha$ is chosen to visualize the
vacuum effect on the decoherence.}}
\end{figure}

We distinguish two different characteristic times of the
decoherence process relative to the vacuum
\begin{equation}\label{tempo di vuoto}
  \tau_{vac}=\frac{1}{\Omega}\;\exp\left[\frac{3\pi}{2\alpha }
  \frac{m_0^2c^2}{(\bbm[p]-\bbm[p]\,')^2}\right] \,,
\end{equation}
and to the thermal contribution
\begin{equation}\label{tempo termico}
   \tau_{th}=\tau_F\frac{3\pi}{2\alpha }
  \frac{m_0^2c^2}{(\bbm[p]-\bbm[p]\,')^2} \,.
\end{equation}
These characteristic times have the same form of those obtained
for the decoherence of the interference pattern in
\cite{Petruccione-Breuer libro 2002}.

The mass and charge parameters $m_0$ and $e$, appearing in
$\tau_{vac}$ and $\tau_{th}$, are arbitrary. The only restriction
is that they refer to a body that can be treated as a point like
particle within the dipole approximation. For example, these
parameters could represent the mass and the charge of a highly
charged nucleus or even of a macroscopic body of linear dimensions
small enough, and therefore $\alpha$ is a free parameter.

Let's observe that the time at which vacuum and thermal
decoherence are effective, depending of the value of the coupling
constant $\alpha$, fall inside the time $\tau_d$ of validity of
our model.

\subsubsection{Analysis of $\Delta p(t)$ and $l_p(t)$}
\label{par:delta p e l di p}

The above results are independent from the structure of the
initial reduced density matrix elements $\rho^{p,p'}_S(0)$. Now we
specialize these results to the case of an initial Gaussian wave
packet of spatial width $\Delta r $
\begin{multline}\label{roiniziale}
\rho^{p,p'}_S(0) = \mathrm{N} \:\exp \left\{
-\frac{3\left[(\bbm[p]-\bbm[p]_0)^2 +
(\bbm[p]\,'-\bbm[p]_0)^2\right]}{4  \Delta p ^2}\right. \\
\left.-i\frac{\bbm[r]_0\cdot
(\bbm[p]-\bbm[p]\,')}{\hslash}\right\} \, ,
\end{multline}
with $\Delta p$ the width in the momentum space, $\bbm[p]_0$ the
initial average momentum of the particle,
$\mathrm{N}=(\sqrt{3}/\sqrt{2\pi} \Delta p)^3$ the normalization
factor and $\Delta r  \Delta p =3 \hslash /2$.
\\
Substituting the gaussian wave packet of Eq.\,(\ref{roiniziale})
in the reduced density matrix at time $t$ of
Eq.\,(\ref{matricemomenti}), this can be put under the form
\begin{flalign}\label{matricemomenti3}
   &\hat{\rho}^{p,p'}_S (t) =\mathrm{N} \exp \left(-\frac{3  p_0^2}{2   \Delta p
  ^2}\right)     \\ & \times
  \exp \left[- \left(\Gamma (t)+\frac{3}{8\Delta p^2}\right) (\bbm[p]-\bbm[p]\,')^2
    -\frac{3(\bbm[p]+\bbm[p]\,')^2}{8\Delta p^2}   \right.  \nonumber   \\ &
  \left. + i \Phi (t)
  (p^2-p'^2)
  +\frac{3\bbm[p]_0\cdot(\bbm[p]+\bbm[p]\,')}{2\Delta p^2}
 -i\frac{\bbm[r]_0\cdot (\bbm[p]-\bbm[p]\,')}{\hslash}  \right]  \nonumber
 .
\end{flalign}
A way to quantify the degree of loss of coherence of the wave
packet is through the coherence length $l_p(t)$ \cite{Libro
decoerenza 2002}, defined as the width of $\hat{\rho}_S(t)$ along
the main skew diagonal, meaning the region inside which the
coherence between momenta has not been yet destructed at time $t$.
$l_p(t)$ may be compared with the width of $\hat{\rho}_S(t)$ along
the diagonal that measures the wave packet width at a time $\Delta
p (t)$, given by
\begin{flalign} \label{larghezzap}
  \Delta  p (t)=\sqrt{<p^2>_t-<p>_t^2}=
   \Delta  p
  \, ,
  \end{flalign}
where we have used $<p^2>=\mathrm{tr}(\hat{\rho}_S(t) p^2)= \Delta
p^2+p_0^2$ and $<\hat{\bbm[p]}>_t=\mathrm{tr}\{\hat{\rho}_S(t)
\hat{\bbm[p]} \}=\bbm[p]_0$. Because $\Delta  p (t)$ is constant
the wave packet doesn't spread with time in momentum space.

The coherence length $l_p(t)$, proportional to the inverse of the
square root of the coefficient of $(\bbm[p]-\bbm[p]\,')^2$ in
Eq.\,(\ref{matricemomenti3}) \cite{Libro decoerenza 2002}, is:
\begin{flalign} \label{lunghezzacoerenzap}
  l_p(t)=
   \frac{\Delta p (t)}{\sqrt{1+8\Delta p^2 \Gamma (t)/3}} \, .
\end{flalign}
To quantify the effective loss of coherence in the wave packet we
study the ratio $ l_p(t)/\Delta p (t)$. This quantity gives a
measure of the relative width of the reduced density matrix off
the diagonal compared with the width along the diagonal. Using
Eq.\,(\ref{lunghezzacoerenzap}) and the explicit form of $\Gamma
(t)$ for $t>\tau_F$, given by Eq.\,(\ref{decoerenza
complessiva2}), we obtain
\begin{flalign} \label{lunghezzacoerenzap2}
  \frac{l_p(t)}{\Delta  p (t)}
  \approx
  \frac{3m_0 c}{4  \Delta p }\sqrt{\frac{\pi \tau_F}
    { \alpha \,t}},\quad\quad  \quad\quad
 t\gg t^*
\end{flalign}
where $t^*= (3m_0c)^2 \pi \tau_F /
 (4\Delta p)^2 \alpha$. Being $\Delta  p (t)$ constant in time,
Eq.\,(\ref{lunghezzacoerenzap2}) shows that the coherence length
for large times decreases going to 0 as $1/\sqrt{t}$ for
$t\rightarrow \infty$. The decoherence process in momentum space
is thus characterized by a complete decay of the off diagonal
elements of the particle density matrix for large times while the
populations remain constant.

This kind of behaviour of the reduced density matrix is shown in
Fig. 3 obtained from Eq.\,(\ref{matricemomenti3}).
\begin{figure}[h]\label{figura1}
\begin{center}
\includegraphics[width=0.35\textwidth, height=0.26\textwidth]{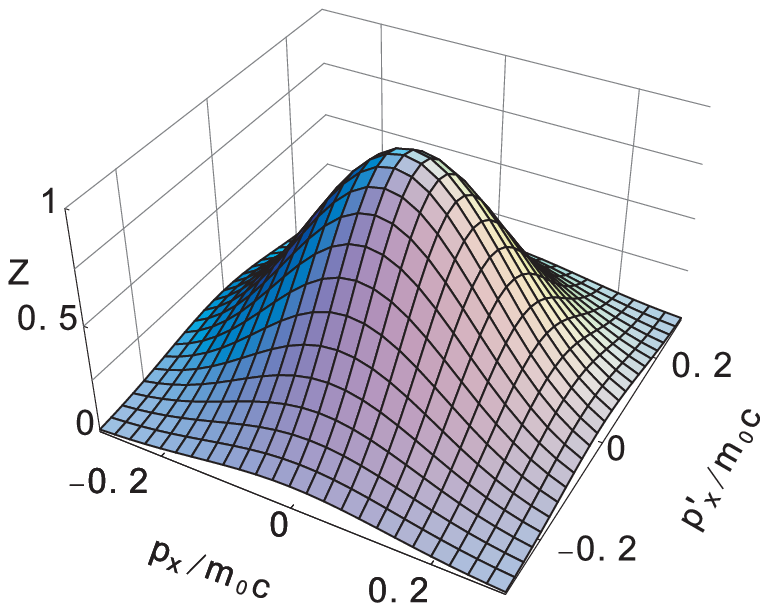}
\includegraphics[width=0.35\textwidth, height=0.26\textwidth]{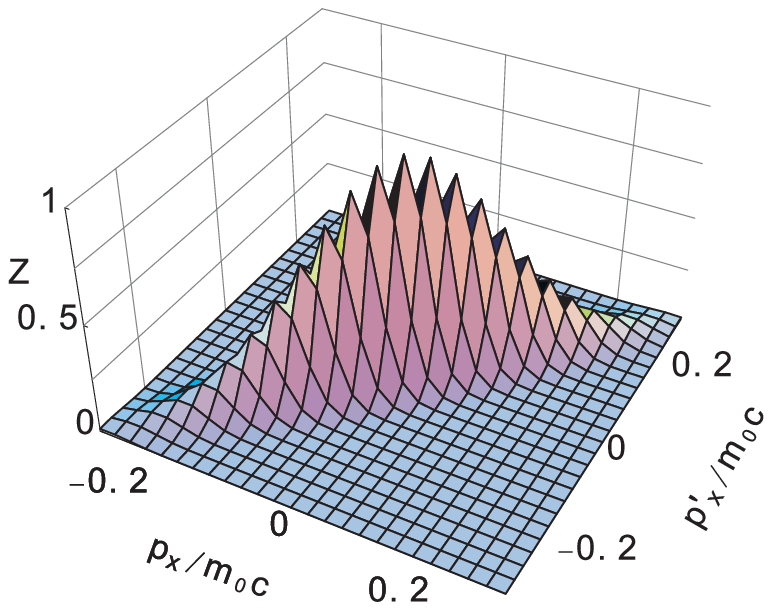}
\caption{\small{In figure it is represented the absolute value of
the normalized density matrix $Z=|\rho^{{p_x},{p'_x}}_S(t)|/N_x$
in one dimension, $N_x=1/\sqrt{2 \pi} \Delta p_x$, with $p_0=0$
and $\Delta p_x/m_0c\approx 0.1$. On the top it is $t=0$ while on
the bottom it is $t=3 \tau_{vac}$, where $\tau_{vac}$ is taken for
$p_x-p_x'=\Delta p_x$.}}
\end{center}\end{figure}

\subsection{Coordinate
space}\label{par:Matrice densita spaziale}

Our analysis is now extended to real space in order to describe
spatial decoherence in more complex situations such as Young
interference or Schr\"{o}dinger cat states setups. We expect that
changing representation the dynamics induced by the interaction
with the electromagnetic field will appear more complex than in
momentum space. As shown, in fact, it provides a basis of pointer
states which allows a simple analysis of the process. To
investigate the effects in the real space we need the reduced
density matrix in the configuration space. It can be obtained from
the corresponding momentum space reduced density matrix by
performing a double Fourier transform:
\begin{flalign} \label{trasformata}
  \rho^{r,r'}_S\!(t)\! =\!\frac{1}{(2\pi\hslash)^3} \!\!\int \!\! \mathrm{d}^3 p \,
  \mathrm{d}^3 p' \rho^{p,p'}_S\!(t) \exp \!\left[ \frac{i}{\hslash}(\bbm[p] \cdot \bbm[r] -
  \bbm[p]\,'\cdot \bbm[r] \, ')\right]\!.
\end{flalign}
Taking the Gaussian wave packet described by $\rho^{p,p'}_S(0)$ of
Eq.\,(\ref{roiniziale}), the transform can be explicitly performed
and is given in Appendix \ref{appendice trasformata}. The spatial
reduced density matrix $\rho^{r,r'}_S(t)$, given by
Eq.\,(\ref{rospaziale2}), can be rewritten as
\begin{flalign}\label{rospaziale4}
  & \rho^{r,r'}_S(t) = \frac{\mathrm{N}\Delta p^3}{\Delta r(t)^3} \:
  \exp\left[i \frac{   \Delta r^2+6\Gamma (t) \hslash^2 }{\Delta
   r(t)^2}
  \frac{ \bbm[p]_0 \cdot
 (\bbm[q]-\bbm[q] \, ') }
  {\hslash }\right]  \nonumber \\
&\times\exp\left\{ - \frac{
3[(\bbm[q]-\langle\bbm[\hat{q}]\rangle_t)^2+
  (\bbm[q] \, '- \langle\bbm[\hat{q}]\rangle_t )^2]}{4\Delta r(t)^2} \right\}
   \\
  &\times \exp\left\{ \frac{ \Delta p^2 }
  { \Delta r(t)^2}\left[-\Gamma (t)(\bbm[q]-\bbm[q] \, ')^2 -i \Phi (t) (q^2-q'^2)\right]
  \right\}
 \nonumber \,,
\end{flalign}
where: $\bbm[q]=\bbm[r]-\bbm[r]_0$ is the displacement from the
initial position and its average at the time $\langle
\bbm[\hat{q}]\rangle_t$ is given by
\begin{equation}\label{mediaposizione}
  \langle\bbm[\hat{q}]\rangle_t=\mathrm{tr}\{\hat{\rho}_S(t)
\hat{q}\}=-2\bbm[p]_0 \Phi (t) \hslash \, ,
\end{equation}
$\Phi (t)$ and $\Gamma (t)$ are defined by Eqs.\,(\ref{decoerenza
complessiva}) and (\ref{Bcompleto}) while $\Delta r(t)$ is the
spatial width of the wave packet at time $t$. \\ From
Eq.\,(\ref{rospaziale2}) we get
\begin{eqnarray}
    \langle\hat{q}^2\rangle_t &=&\mathrm{tr}\{\hat{\rho}_S (t) \hat{q}^2\}
    \\ &=&\Delta
r^2+6\Gamma (t)\hslash^2+\frac{9\Phi (t)^2\hslash^4}{\Delta
r^2}+4p_0^2\Phi (t)^2\hslash^2  \nonumber\, ,
\end{eqnarray}
and thus, using also Eq.\,(\ref{mediaposizione}), $\Delta r(t)$ is
given by:
\begin{flalign} \label{lunghezzacoerenzar}
  \Delta r(t)&= \Delta q(t)=\sqrt{\langle\hat{q}^2\rangle_t-\langle\bbm[\hat{q}]\rangle_t^2} \nonumber \\& =
   \Delta r \sqrt{1 + \frac{6 \Gamma (t) \hslash^2}{\Delta r^2} + \frac{9 \Phi (t)^2 \hslash^4}{\Delta r^4}}
   \,.
\end{flalign}
$\Delta r(t)$ is at $t=0$, $\Delta r(0)=\Delta r$, and increases
with time.

$\rho^{r,r'}_S(t)$ can be obtained by its form at $t=0$
\begin{multline}\label{rospazialeiniziale}
  \rho^{r,r'}_S \!(0)=\frac{\mathrm{N}\Delta p^3}{\Delta r^3} \:
  \exp \! \left[i \frac{ \bbm[p]_0 \cdot
 (\bbm[q]-\bbm[q] \, ') }{\hslash}
 - \frac{  3(q^2+  q'^2)}{4\Delta r^2} \right] ,
\end{multline}
replacing the initial width of the wave packet $\Delta r$ with its
value at time $t$, $\Delta r(t)$, multiplying  by $ (\Delta
r^2+6\Gamma (t) \hslash^2) /\Delta r(t)^2$ the phase factor in the
first exponent, centering the wave packet in the average
displacement $\langle\bbm[\hat{q}]\rangle_t$ in the second
exponent and multiplying by an exponential factor which gives an
increase of decoherence and a phase variation analogous to the
factor appearing in the reduced momentum density matrix elements
of Eq.\,(\ref{matricemomenti}).

\subsubsection{Time dependent dressing}\label{par:variazione di
massa}

The average of the operator $\hat{\bbm[q]}$ at time $t$, given by
Eq.\,(\ref{mediaposizione}) and using the explicit form of $\Phi
(t)$ (\ref{Bcompleto}), is
\begin{flalign}\label{mediaposizione2}
  \langle\bbm[\hat{q}]\rangle_t=
  \frac{\bbm[p]_0 t}{m_0}\left[1-\frac{4\alpha\hslash\left(\Omega-\frac{\arctan \Omega
  t}{t}\right)}
  {3\pi  m_0 c^2} \right]\,.
\end{flalign}
From this equation the average velocity of the wave packet is
\begin{flalign}\label{media velocita}
  \langle\hat{\dot{\bbm[q]}}\rangle_t=\frac{\mathrm{d}}{\mathrm{d} t}\langle\bbm[\hat{q}]\rangle_t=\frac{\bbm[p]_0 }
  {m_0}\left(1-\frac{4\alpha\hslash \Omega}{3\pi m_0 c^2}
  \frac{\Omega^2t^2}{1+\Omega^2t^2}\right) \,.
\end{flalign}
As observed before, $\hat{\bbm[p]}$ is a constant of motion,
instead the velocity $\hat{\dot{\bbm[q]}}=[\hat{\bbm[p]} - e
\bbm[\hat{A}] (\bbm[r]_0)/c]/m_0$ it is not because it does not
commute with the Hamiltonian (\ref{hamiltoniana}). This may be
related to the fact that, starting with uncoupled initial
conditions, the charged particle is subject to time dependent
dressing by the transverse photons. This increases its mass while
$\hat{\bbm[p]}$ remains constant. The mass variation can be
obtained casting Eq.\,(\ref{media velocita}) in the form
\begin{flalign}\label{velocita media al tempo t}
 \langle\hat{\dot{\bbm[q]}}\rangle_t= \frac{\langle\bbm[\hat{p}]\rangle_t}{m (t)}\approx
  \frac{\bbm[p]_0 }{m_0}\left[1-\frac{\delta m (t)}{m_0}\right]
   \, ,
\end{flalign}
where $m(t)=m_0+\delta m (t)$ is the mass at time $t$ being the
mass increase $\delta m(t)$ given by
\begin{equation} \label{variazione di massa}
 \delta m(t)=\frac{4\alpha\hslash \Omega}{3\pi c^2}
  \frac{\Omega^2t^2}{1+\Omega^2t^2} \:\approx\:\left\{\:\begin{aligned}
\frac{4\alpha\hslash\Omega}
  { 3\,\pi \,c^2 }\Omega^2 t^2, \quad\quad \, \textit{t}\ll\Omega^{-1}
  \\ \\
\frac{4\,\alpha\,\hslash\,\Omega}
  {3\,\pi \,c^2 }\;.\;\;\;\;\; \quad\quad  \textit{t}\gg\Omega^{-1}
\end{aligned}
\right.
\end{equation}
For $t\ll\Omega^{-1}$ $\delta m(t)$ increases quadratically
\cite{bellomo 2004} while for $t\gg\Omega^{-1}$ coincides with the
usual total mass variation due to the interaction with the
electromagnetic field \cite{Sakurai 1977}.

We observe that the equation of motion (\ref{mediaposizione2}),
from which we derived the expression for the mass increase, is
related only to the total phase factor $\Phi(t)$ and is then
temperature independent at first order in $v_0/c$.

\subsubsection{Analysis of $\Delta r(t)$ and $l_r(t)$}
\label{par:delta r e l di r}

The mass variation due to dressing is relevant if one wishes to
compare the evolution of the wave packet width in the absence of
interaction, $\Delta r(t)^0$, with its expression, $\Delta r(t)$,
in the presence of interaction. In the last case we have from
Eq.\,(\ref{lunghezzacoerenzar})
\begin{flalign}\label{lunghezzacoerenzar2}
  \Delta r(t)&=
  \Delta r\sqrt{1 + \frac{\Delta p^2 t^2}{\Delta r^2}\frac{4\hslash^2\Phi (t)^2}{t^2}
  +\frac{6 \Gamma (t) \hslash^2}{\Delta r^2}} \,,
\end{flalign}
with $\Gamma (t)$ and $\Phi (t)$ defined by Eqs.\,(\ref{decoerenza
complessiva}) and (\ref{Bcompleto}). Putting $e=0$ we obtain the
well known expression for the free spread \cite{Pauli 2000}
\begin{equation}\label{evoluzionelibera}
  \Delta r(t)^0=\Delta r \sqrt{1+
\frac{\Delta p^2 t^2}{\Delta r^2}\frac{1}{m_0^2}} \,.
\end{equation}
$\langle\bbm[\hat{q}]\rangle_t$, given by
Eq.\,(\ref{mediaposizione}), can also be obtained by integrating
Eq.\,(\ref{velocita media al tempo t})
\begin{equation}\label{mediaposizione0}
  \langle\bbm[\hat{q}]\rangle_t=-2\bbm[p]_0 \hslash \Phi (t)
  =\bbm[p]_0\int_0^t\frac{1}{m(t')}\mathrm{d}t'\,.
\end{equation}
Thus we can identify
\begin{equation}\label{B rispetto alla massa}
  \frac{-2\hslash \Phi (t)}{t}=\frac{1}{t}\int_0^t\frac{1}{m(t')}\mathrm{d}t'
={\left\langle \frac{1}{m(t')}\right\rangle}_t \, ,
\end{equation}
${\left\langle \frac{1}{m(t')}\right\rangle}_t$ being the time
average of $1/m(t')$ over the time $t$. The width of the wave
packet at time $t$ (\ref{lunghezzacoerenzar2}) can be thus
rewritten as
\begin{flalign}\label{spread totale}
  \Delta r(t)=
  \Delta r\sqrt{1 + \frac{\Delta p^2 t^2}{\Delta r^2}{\left\langle \frac{1}{m(t')}\right\rangle}_t^2
   +\frac{6 \Gamma (t)  \hslash^2}{\Delta r^2}}  \,.
\end{flalign}
Eq.\,(\ref{spread totale}) shows that, starting from uncoupled
condition, the interaction with the electromagnetic field induces
differences with respect to the free evolution $\Delta r(t)^0$.
The first one consists in the replacement of the inverse of the
initial mass $1/m_0$ by ${\left\langle 1/m(t')\right\rangle}_t$
and may be attributed to the $t$-dependent dressing. This effect
is due to the vacuum fluctuations and is related to the total
phase factor $\Phi(t)$, the  mass increase leading to a rate
decrease of the width with respect to the free case. The second
effect is given by the term within the square root
\begin{equation}\label{spread dovuto a decoerenza}
  \frac{6 \Gamma (t)\hslash^2 }{\Delta r^2}=\frac{1}{\Delta r^2} \frac{4
\alpha \hslash^2 }{ \pi m_0^2 c^2 } \ln \left[ \sqrt{1 +
\Omega^2t^2} \, \, \frac{\sinh(t/\tau_F)}
  {t/\tau_F}  \right] \, .
\end{equation}
It always leads to an additional increase of the width of the wave
packet. It contains both the effect of vacuum, represented by the
term $\sqrt{1 + \Omega^2t^2}$, and of the thermal field
represented by the term $\sinh(t/\tau_F)/ (t/\tau_F)$, being this
last term for $T=0$ equal to $1$ ($\tau_F=\hslash/\pi k_BT$).

The comparison of the amplitudes of the vacuum and thermal terms
in time may be obtained using the forms of the coefficients
$\Gamma (t)$ and $\Phi (t)$ given by Eqs.\,(\ref{decoerenza
complessiva2}) and (\ref{fattore di fase approssimazioni}) for
small ($t\ll \Omega^{-1}$) and large ($t\gg \tau_F$) times . For
small times the total effect is that the width of the wave packet
results larger than in the free case. For large times, instead,
the additional term becomes negligible and  the spreading is
slower than in the free case because the increasing of mass.

The space coherence length $l_r(t)$ represents the typical
distance for which it is possible to have constructive
interference among different parts within the wave packet. It can
be read directly from the coefficient of $(\bbm[q]-\bbm[q]\,')^2$
term of the reduced density matrix written under the form of
Eq.\,(\ref{rospaziale3}), being in fact proportional to the
inverse of this coefficient \cite{Libro decoerenza 2002}:
\begin{flalign}\label{lunghezza di coerenza spaziale}
l_r(t) =\frac{\Delta r(t)}{\sqrt{1 +6 \hslash^2 \Gamma (t)/\Delta
r^2}}.
\end{flalign}
Using Eq.\,(\ref{lunghezzacoerenzar}) for $\Delta r(t)$ it results
that $l_r(t)$ increases with time, while, analogously to what
happens in momentum space (\ref{lunghezzacoerenzap}), $
l_r(t)/\Delta r(t)$ decreases with time because $\Gamma (t)$
increases with time (\ref{decoerenza complessiva}). In absence of
interaction $\Gamma (t)$ is equal to zero and therefore the free
space coherence length, $l_r(t)^0$, is always equal to the width
of the wave packet which increases coherently in time due to the
well known free spread (\ref{evoluzionelibera}). The coupling with
the field induces an evolution of $l_r(t)$ different from $\Delta
r (t)$. Using Eqs.\,(\ref{lunghezzacoerenzap}) and (\ref{lunghezza
di coerenza spaziale}) and $\Delta  r \Delta p=3\hslash/2$, it
follows that $l_r(t)/\Delta r(t)=l_p(t)/\Delta p(t)$ and therefore
Eq.\,(\ref{lunghezzacoerenzap2}) describes also in the coordinate
space the behaviour of the coherence length with respect to the
width of the wave packet for large times. This equation shows that
the ratio decreases to zero as $1/\sqrt{t}$ for $t\rightarrow
\infty$ describing a loss of coherence also in the configuration
space.

Another interesting aspect to investigate is the behaviour of
$l_r(t)$ with respect to its evolution in the free case
$l_r(t)^0$. Using Eq.\,(\ref{evoluzionelibera}) for $l_r(t)^0$ and
Eqs.\,(\ref{lunghezza di coerenza spaziale}) and (\ref{spread
totale}) we can put the coherence length in the form:
\begin{flalign}\label{confronto lunghezze di coerenza spaziali}
  l_r^2(t)&= \Delta r^2 +\Delta p^2 t^2{\left\langle
  \frac{1}{m(t')}\right\rangle}_t^2 \frac{\Delta r^2}{\Delta r^2 + 6 \Gamma (t)  \hslash^2
  }<  l_r^2(t)^0  \,.
\end{flalign}
Eq.\,(\ref{confronto lunghezze di coerenza spaziali}) shows that
dressing induces a slower increase of coherence length due to the
mass increase, but always maintaining the coherence, while vacuum
and thermal field induce a destruction of coherence in space such
that the coherence length is lower than in the free evolution
case.

In the momentum space we obtained a simple dynamics: the width of
the wave packet remains constant while the coherence length
decreases with respect to its initial value going to zero. In
coordinate space, instead, different factors contribute to the
dynamics: free evolution contributes to the coherent increase of
the width of the wave packet coherently and therefore of the
coherence length; the particle time dependent dressing of the
particle slows this increase; finally vacuum and thermal field
induce a loss of space coherence such that the value of the space
coherence length in presence of the interaction is always lower
than its value in absence of the interaction.

\subsubsection{Linear entropy}\label{par:Entropia lineare}

The dynamics of our system is described by the reduced density
matrix time evolution as a transformation from the pure initial
state (\ref{stato iniziale}) into a statistical mixture
(\ref{matricemomenti}). The time dependence of this process, that
implies a loss of information on the system, may be described by
the so-called linear entropy, $S_{\mathrm{lin}}$
\cite{Petruccione-Breuer libro 2002}. It has been analyzed in the
case of localization by scattering, to measure how strongly the
environment destroys coherence between positions by delocalizing
phases, finding a linear departure in time from the initial value
$0$ describing a pure state \cite{Libro decoerenza 2002}. Using
its definition we obtain here
\begin{equation}\label{entropia lineare}
  S_{\mathrm{lin}}=\mathrm{tr}(\hat{\rho}_S-\hat{\rho}_S^2)=1-\sqrt{\frac{1}{1+6 \Gamma (t) \hslash^2
  /\Delta r ^2}}\,,
\end{equation}
which describes the loss of purity of the initial state. It is
interesting to note that in the case of initial Gaussian wave
packet \cite{Morikawa 1990}, $S_{\mathrm{lin}}$ is directly
connected to a dimensionless measurement of the decoherence given
by the ratio between the decoherence length and the wave packet
width. This ratio coincides both in the $p$ and $r$
representations (\ref{lunghezzacoerenzap}) and (\ref{lunghezza di
coerenza spaziale}) and using Eq.\,(\ref{entropia lineare}) may be
expressed as
\begin{equation}\label{decoerenzaadimensionale}
  S_{\mathrm{lin}}= 1-\frac{l_p(t)}{\Delta p (t)}=1-
  \frac{l_r(t)}{\Delta r(t)}\,.
\end{equation}
Using Eq.\,(\ref{lunghezzacoerenzap}) and the approximated form of
$\Gamma(t)$ for small times given by Eq.\,(\ref{decoerenza
complessiva2}), we find that $S_{lin}$ at the beginning evolves
quadratically from the initial value $0$ corresponding to a pure
state, then slows and finally (\ref{lunghezzacoerenzap2}) goes to
1 for $t\rightarrow\infty$ as $1- 1/\sqrt{t}$.

In Fig. 4 the time development of the linear entropy is plotted as
a function of the coupling constat $\alpha$. The figure shows
clearly that the increase of this quantity towards 1 depends
strongly on $\alpha$, that is on the charge of the particle
considered.
\begin{figure}[h]\label{linear entropy} \centering
\includegraphics[width=0.45\textwidth, height=0.33\textwidth]{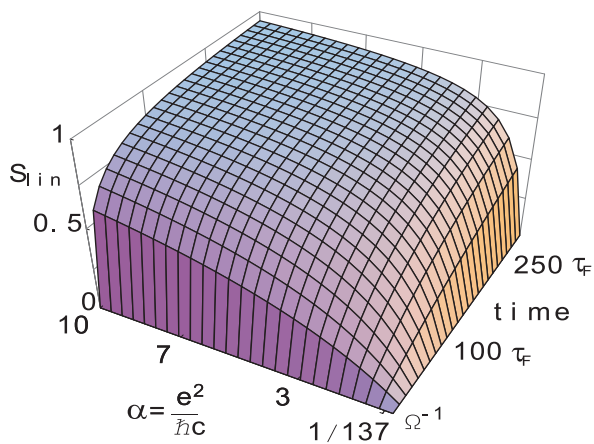}
\caption{\small{In figure it is shown the behaviour of
$S_{\mathrm{lin}\mathrm{}}$ in time as a function of $\alpha$.}}
\end{figure}

\section{Interpretation of vacuum induced decoherence }\label{par:vacuum contribution interpretation}

The temperature independent part of decoherence is represented by
$\Gamma^{p,p'}_{vac}(t)$ of Eq.\,(\ref{contributo di vuoto}). In
the following we shall analyze the processes that contribute to
$\Gamma^{p,p'}_{vac}(t)$.

In the case of a charged particle, initially described by a wave
function made of a coherent superposition of two moving wave
packets, it has been previously shown \cite{Breuer-Petruccione
2001} that Bremsstrahlung radiation induces decoherence decreasing
the visibility of the interference pattern that results from their
overlapping. The reason is that in its trajectory the particle is
subject to a sudden change of the 4-momentum and in this process
it is radiated as Bremsstrahlung photons the energy
\cite{Peskin-Schroder 1995}:
\begin{equation}\label{energia di bremsstrahlung}
  E_R=\frac{\alpha}{\pi} k_{max} I(v,v')\, ,
\end{equation}
where $I(v,v')$ is a function of the initial and final velocity
and $k_{max}$ is the wave vector corresponding to the frequency
equals to the reciprocal of the time scattering during which the
4-momentum changes. This energy results, as also the decoherence
function, proportional to $\alpha$. Thus Bremsstrahlung may be
hold responsible of decoherence.

In our system the particle is also subject to a change of velocity
during the dressing process with the emission of Bremsstrahlung
photons. These could be held responsible of the temperature
independent loss of coherence between the momentum components of
the wave packet. However the radiation energy emitted in the unity
of time from the accelerated charged particle during the dressing
process can be estimated as \cite{Rossi 1991}:
\begin{equation}\label{flusso di energia}
  E_B \propto \frac{e^2 \langle\hat{\ddot{\bbm[q]}}\rangle^2}{c^3}=\alpha \frac{ \hslash   \langle\hat{\ddot{\bbm[q]}}\rangle^2}{c^2}
  \,,
\end{equation}
with $\langle\hat{\ddot{\bbm[q]}}\rangle$ being the average
acceleration of the particle.  To obtain
$\langle\hat{\ddot{\bbm[q]}}\rangle$ during the dressing we take
the time derivative of Eq.\,(\ref{media velocita}):
\begin{equation}\label{accelerazione}
  \langle\hat{\ddot{\bbm[q]}}\rangle=-\frac{\bbm[p]_0 }{m_0^2}\frac{4\,\alpha\,\hslash\,\Omega}
  {3\,\pi \,c^2 } \frac{2\Omega^2 t}{\left(1+\Omega^2t^2\right)^2}
  \,.
\end{equation}
Substituting this last equation in Eq.\,(\ref{flusso di energia})
the estimated emitted energy per unity of time results
proportional to $\alpha^3$. The vacuum contribution to the
decoherence function is shown from Eq.\,(\ref{contributo di
vuoto}) to be proportional to $\alpha$. From the considerations
above it follows that the emission of Bremsstrahlung photons
doesn't seem to be relevant for the vacuum decoherence process.

However let's observe that for short times ($t\ll \Omega^{-1}$),
the decoherence factor $\Gamma(t)$ of Eq.\,(\ref{decoerenza
complessiva2}) and the mass variation $\delta m (t)$ of
Eq.\,(\ref{variazione di massa}) show both the same $t$ and
$\alpha$ dependence. This appears to suggest a connection between
the decoherence process for small times (vacuum contribution) and
the dressing process. In analogy to the case of the two level
systems \cite{Palma-Suominen-Ekert 1996}, the link between
dressing and vacuum induced decoherence could be attributed to the
correlation that get established between each component $\bbm[p]$
of the wave packet and the part of the dressing structure of the
transverse electromagnetic field associated to it.

To verify this hypothesis, we shall analyze the evolution of the
field associated to each component of the wave packet, during the
initial phase of the decoherence process.

\subsection{Field structure dynamics}\label{par:Cosa succede al
campo?}

In the analyses of decoherence the behaviour of the environment is
usually not investigated being the interest placed on the system
evolution. In our case the environment is the electromagnetic
field and its behaviour during the decoherence process can be
analyzed by performing the trace of the total density matrix over
the degrees of freedom of the particle.

For calculation purposes we shall consider the initial wave packet
of momentum width $\Delta p$ as a sum of momentum sharp wave
packets of width $\Delta \bar{p} \ll \Delta p$. Each of these
sharp wave packets is centered at a momentum $\bar{\bbm[p]}$ and
it has in configuration space a width $\Delta \bar{r} \approx
\hslash/\Delta \bar{p}$ taken less than $c/\Omega$ so that the
dipole approximation can be yet used. To describe the development
of the field correlated to one of these sharp wave packets
centered at $\bar{\bbm[p]}$ we start from a totally decoupled
initial condition. The field is taken in its vacuum state and the
charged particle is described by a sharp wave packet with momentum
components peaked around $\bar{\bbm[p]}$ of the form $
\sqrt{N_\varepsilon} \sum_p
\delta^\varepsilon_{p,\bar{p}}|p\rangle$, where $N_\varepsilon$ is
a normalization factor and $\delta^\varepsilon_{p,\bar{p}}$
indicates a quasi delta centered on $\bar{\bbm[p]}$ of width
$\Delta \bar{p}$.

The corresponding initial density matrix is
\begin{flalign}\label{stato iniziale campo}
  \hat{\rho} (0)&=\hat{\rho}_S (0)\otimes |\{0_{\bar{k},\bar{j}}\}\rangle \langle
  \{0_{\dot{k},\dot{j}}\}| \nonumber \\ & =\sum_{p,p'}|p\rangle \bar{N }\delta^\varepsilon_{p,\bar{p}}
  \delta ^{\varepsilon '}_{p',\bar{p}}
  \langle p'|  \otimes   |\{0_{\bar{k},\bar{j}}\}\rangle \langle
  \{0_{\dot{k},\dot{j}}\}| \, .
\end{flalign}
We shall consider the representation of $\hat{\rho}(t)$ in a
coherent basis. Indicating with $ |\lambda_{k,j} \rangle $ a
coherent state of the mode $\{\bbm[k],j \}$ of amplitude
$\lambda$, the reduced density matrix elements of the field in
this basis, with the initial condition of Eq.\,(\ref{stato
iniziale campo}), are given by
\begin{multline}\label{elementi matrice ridotta campo}
  \hat{\rho}_F^{\lambda_{k,j},\lambda'_{k',j'}}(t)=\langle \lambda_{k,j}|\mathrm{tr}_S \{\hat{U}(t)
   \hat{\rho}_S (0)\otimes
 |\{0_{\bar{k},\bar{j}}\}\rangle  \\ \langle \{0_{\dot{k},\dot{j}}\}
  | \hat{U}^{-1}(t)\}|\lambda'_{k',j'}\rangle \,.
\end{multline}
The explicit calculation, reported in Appendix \ref{dinamica
campo} by Eq.\,(\ref{elementi matrice ridotta campo3}), gives for
the reduced density matrix of the field
\begin{multline}\label{elementi matrice ridotta campo4}
  \hat{\rho}_F^{\lambda_{k,j},\lambda'_{k',j'}}(t)=N_\varepsilon  \exp \left[-\frac{|\lambda_{k,j}|^2}{2}
  -\frac{|\lambda'_{k',j'}|^2}{2}-\frac{|\beta_{k,j}^{\bar{p}}|^2}{2}   \right. \\  \left.
  -\frac{|\beta_{k',j'}^{\bar{p}}|^2}{2}
  +\lambda^{*}_{k,j}\beta_{k,j}^{\bar{p}}
  +\lambda'_{k',j'}\beta_{k',j'}^{{\bar{p}}*} \right] \,,
\end{multline}
with $\beta_{k,j}^{\bar{p}}$ defined in Eq.\,(\ref{beta}). Because
of our choice of sharp wave packets in $p$ space, the density
matrix of Eq.\,(\ref{elementi matrice ridotta campo4}) retains
only a dependence on $\bar{p}$.

Eq.\,(\ref{elementi matrice ridotta campo4}) allows to get the
average number of photons $\langle \hat{n}_{\bar{p}}\rangle$ that
can be associated to each sharp wave packet of width $\Delta
\bar{p}$ and centered at the momentum $\bar{\bbm[p]}$ of the total
wave packet. The calculation, performed in Appendix \ref{dinamica
campo} by (\ref{media fotoni5}), leads to
\begin{flalign}\label{media fotoni6}
  \langle \hat{n}_{\bar{p}}\rangle=
  \frac{2 \alpha }{3\pi }\frac{{\bar{p}}^2}{m_0^2 c^2}\ln
  \left(1+\Omega^2t^2\right) \,.
\end{flalign}
The time dependence of the average number of photons of
Eq.\,(\ref{media fotoni6}) is, apart a factor 2, equal to that of
the vacuum contribution to the decoherence function
(\ref{contributo di vuoto}). This result appears to give a strong
indication that it is just the buildup of correlations among the
various momenta that compose the wave packet and the corresponding
associated transverse photons that leads to vacuum decoherence in
our system.

To confirm the possibility of associating a number of photons to
the various momentum components of a given wave packet, we could
choose as initial state a sum of two sharp wave packets of width
$\Delta \bar{p}$ peaked around two different momenta. In this case
it is easy to show that the average number of photons surrounding
the particle can be written as a sum of two terms relative to the
two sharp wave packets composing the initial state.

The energy associated to the field structure that builds up around
the particle is responsible together with the interaction energy
of the mass variation $\delta m$ computed in Eq.\,(\ref{variazione
di massa}). The average energy associated to the cloud of photons,
obtained in Appendix \ref{dinamica campo} by (\ref{energia media
fotoni}), is equal to
\begin{flalign}\label{energia media fotoni2}
  \langle \hat{E}_{F}\rangle  =
  \frac{8\alpha}{3\pi}\frac{\hslash \Omega }{m_0 c^2}
  \frac{\Omega^2t^2}{1+\Omega^2t^2}\frac{{\bar{p}}^2}{2m_0} \,.
\end{flalign}
$\langle \hat{E}_{F}\rangle$ can be written, using
Eq.\,(\ref{variazione di massa}) for $\delta m$, as $\langle
\hat{E}_F \rangle=-({\bar{p}}^2/2m_0)\times (\delta_F m/m_0) $
with $\delta_F m=-2\delta m$. Therefore $\langle
\hat{E}_{F}\rangle$ reflects on one side the build up of
correlations with momenta and on the other side contributes to the
mass variation. This explains the analogous time behaviour of
$\delta m (t)$ and $\Gamma_{vac} (t)$.

\section{Summary and Conclusions}\label{par:Summary and Conclusions}

We have considered a free charged particle interacting with a bath
consisting of an electromagnetic field at temperature $T$. We have
analyzed the decoherence on the charged particle wave packet
induced by the interaction through the investigation of the off
diagonal elements of the particle reduced density matrix. The
interaction has been taken in the minimal coupling form and the
particle is described by a wave packet of width $\Delta r$. The
effect of all the modes of wavelength larger than $\Delta r$ can
be taken into account within the dipole approximation. The dipole
approximation and the neglecting of the quadratic potential term
reduces the coupling to a linear form and this in turn allows an
exact treatment of the dynamics of the system.

Our analysis has been conducted in the context of non relativistic
QED which is in the spirit of modern quantum field theory an
effective low energy theory with the cut off frequency $\Omega$
parameterizing the physics due to the higher frequencies \cite{Zee
2003}. For this reason our final results must show a dependence on
$\Omega$, that is however as usual weak (logarithmic), as for
example in the case of non relativistic expression for the Lamb
shift.

The analysis of the decoherence process has been conducted both in
the momentum and configuration space and it has been possible to
separate both the vacuum and the thermal contribution to
decoherence.

In momentum space decoherence among different momentum components
occurs without population decay, therefore decoherence occurs in
its purest form that is without dissipation. This is reflected by
the fact that the width $\Delta p (t)$ of the wave packet remains
constant in time while the coherence length $l_p (t)$ decreases in
time, in particular as $1/\sqrt{t}$ for large $t$.

In configuration space again both vacuum and thermal contribution
appear in the decay of the off diagonal elements of the reduced
density matrix similarly to what occurs in the momentum space.
However in the characterization of the development of decoherence
by the behaviour of the space width of the wave packet, $\Delta r
(t)$, and the coherence length, $l_r(t)$, it is necessary to
consider that in these quantities two contributions appear, which
are not present in the momentum space. The first is due to the
free evolution of the wave packet and the second to the dressing
process. The appearance of these contributions only in the
configuration space is due to the fact that the Hamiltonian
commutes with each momentum component that then results to be a
constant of the motion. In particular the dressing process, with
the emission and absorption of virtual photons and the creation of
a structure of transverse field around the particle, doesn't
modify the distribution of momenta of the wave packet while it
modifies the spatial probability distribution. We have determined
the contribution of these physical effects to $\Delta r (t)$ and
$l_r(t)$.

We have tried to determine the physical effect responsible for the
part of decoherence independent from the temperature. The
Bremsstrahlung photons emitted during the dressing have been shown
not to be relevant for vacuum decoherence. The results obtained
about the particle mass variation indicate that the vacuum
contribution to decoherence is temporally linked to the dressing
process. We have shown by the analysis of the field structure
dynamics that the onset of time dependent correlations, induced by
the interaction, between the momentum components of the particle
wave packet and the associated field structure, may be held
responsible of vacuum induced decoherence. In fact the average
number of entangled photons with a given momentum has the same
time dependence of the vacuum part of the decoherence function and
moreover has the same dependence on the physical parameters of the
system.

The results obtained for our system on the development of induced
decoherence depend on the fact that in the initial state
considered there are not particle-field correlations. Previously
it has been shown that decoherence evolution is influenced by the
presence of initial partial correlation \cite{bellomo2
2005,Smith-Caldeira 1990, Romero-Paz 1997,Lutz 2003}. It appears
of interest to analyze in which way the results obtained in this
paper are modified in the more realistic case in which partial
correlations between the system and the environment are present
since the beginning.

\appendix
\section{} \label{funzione di decoerenza calcolo}
\subsection{Phase factor}
In order to compute the phase factor $\xi(p,t)$ of
Eq.\,(\ref{operatore evoluzione cbh}) it is necessary to explicit
the commutator of $\hat{H}_I$ at different times
(\ref{commutatore}). Using the following relation satisfied by the
polarization vectors
\begin{equation}\label{somma polarizzazioni}
  \sum_j \varepsilon_{k,j}^m
\varepsilon_{k,j}^n=\delta_{m,n}
-\check{\bbm[k]}_m\check{\bbm[k]}_n \,,
\end{equation}
where $\check{\bbm[k]}$ indicates the versor of $\bbm[k]$ and $m$
and $n$ are generical components, we obtain
\begin{equation} \label{p scalare epsilon}
  \sum_{j}|\bbm[p]\cdot
  \bbm[\varepsilon]_{k,j}|^2= p^2-(\bbm[p] \cdot
  \check{\bbm[k]})^2 \,.
\end{equation}
By using the explicit form of the coupling coefficients of
Eq.\,(\ref{coefficiente di accoppiamento}), and Eq.\,(\ref{p
scalare epsilon}) to compute the sum over the polarizations, the
commutator of $\hat{H}_I$ at different times (\ref{commutatore})
assumes the form
\begin{flalign}\label{commutatore2}
&[\hat{H}_I(s),\hat{H}_I(s')]\!=\!-2i \! \sum_{p,k,j} \!
g_{k,j}^{p\,2} \hat{\sigma}_p \sin
\left[(\omega_k-\bbm[k]\cdot\bbm[v]_0 )(s-s')\right] \nonumber \\
  &= \! -i\frac{4 \pi e^2 \hslash }{m_0^2 V} \! \sum_{p,k} \! \frac{p^2-
  (\bbm[p] \cdot
  \check{\bbm[k]})^2}{\omega_k}\hat{\sigma}_p \sin\left[(\omega_k-\bbm[k]\cdot\bbm[v]_0 ) (s-s')\right] .
\end{flalign}
The time integrations present in Eq.\,(\ref{operatore evoluzione
cbh}) give
\begin{multline} \label{integrale su tempi}
  \int_0^t \mathrm{d}s \int_0^t
  \mathrm{d}s' \sin \left[(\omega_k-\bbm[k]\cdot\bbm[v]_0 )(s-s')\right] \theta (s-s') \\
  =\frac{1}{(\omega_k-\bbm[k]\cdot\bbm[v]_0 )}\left[ t-\frac{\sin(\omega_k-\bbm[k]\cdot\bbm[v]_0 ) t}{(\omega_k-\bbm[k]\cdot\bbm[v]_0 )
  }\right]\,,
\end{multline}
and joining Eqs.\,(\ref{integrale su tempi}), (\ref{commutatore2})
and (\ref{operatore evoluzione cbh}) we obtain for $\xi(p,t)$
\begin{multline} \label{fase prima continuo}
  \xi(p,t)=\frac{2\pi\,e^2  }{m_0^2 V \hslash} \sum_{k} \frac{p^2-(\bbm[p] \cdot \check{\bbm[k]})^2}
  {\omega_k (\omega_k-\bbm[k]\cdot\bbm[v]_0 )}\\
\times \left[ t-\frac{\sin(\omega_k-\bbm[k]\cdot\bbm[v]_0 )
t}{(\omega_k-\bbm[k]\cdot\bbm[v]_0 )
  }\right] \,.
\end{multline}
By taking the continuum limit on the field modes, $\sum_k
\rightarrow V\int_0^\infty \mathrm{d}^3 k / (2\pi )^3$,
Eq.\,(\ref{fase prima continuo}) becomes
\begin{multline}\label{fase continuo} \xi(p,t)=\frac{ e^2
}{\hslash m_0^2 (2\pi)^2 } \int \mathrm{d}^3 k \left[
t-\frac{\sin(\omega-\bbm[k]\cdot\bbm[v]_0 )
t}{(\omega-\bbm[k]\cdot\bbm[v]_0 )
  }\right]
   \\
  \times \frac{p_x^2(1-\check{\bbm[k]}^2_x)+p_y^2(1-\check{\bbm[k]}^2_y)+p_z^2(1-\check{\bbm[k]}^2_z)}
  {\omega (\omega-\bbm[k]\cdot\bbm[v]_0 )} \,.
\end{multline}
By introducing the cut off factor $\exp (-\omega/\Omega)$,
Eq.\,(\ref{fase continuo}) assumes the form
\begin{flalign}\label{fase continuo2}
\xi(p,t)=&\frac{ e^2 }{4 \pi^2 \hslash m_0^2 c^3 }\int_0 ^\infty
\frac{\mathrm{d}\omega}{\omega}
\exp\left(-\frac{\omega}{\Omega}\right)   \\
  & \times \int
\mathrm{d}o  \frac{\omega t(1-X)-\sin \left[\omega
t(1-X)\right]}{{(1-X)^2} }
  \nonumber \\
  & \times
  \left[p_x^2(1-\check{\bbm[k]}^2_x)+p_y^2(1-\check{\bbm[k]}^2_y)+p_z^2(1-\check{\bbm[k]}^2_z)\right] \nonumber
  \,,
\end{flalign}
where $\mathrm{d}o =\sin \theta \mathrm{d}\theta \mathrm{d}\varphi
$ is the infinitesimal solid angle and we have posed
\begin{multline}\label{kappavizero}
\bbm[k] \cdot \bbm[v]_0=\omega \frac{v_0}{c} \left[ \sin \theta_0
\sin \theta_k \cos (\varphi_k-\varphi_0)+\cos
\theta_0\cos\theta_k\right] \\ =\omega X \,,
\end{multline}
where $\theta_0$ and $\varphi_0$ are the angles of the vector
$\bbm[v]_0$ and $\theta_k$ and $\varphi_k$ are the angles of the
vector $\bbm[k]$.
\\
Indicating with $f(X)$ the $X$ dependent part within the integrand
in Eq.\,(\ref{fase continuo2}), for small values of $v_0/c$, this
can be expanded with respect to $X$ obtaining up to the first
order in $X$
\begin{multline}\label{espansion1}
 f(X)= \frac{\omega t(1-X)-\sin \left[\omega
t(1-X)\right]}{{(1-X)^2} } \\ \approx \omega t-\sin \omega t
+f'(0)X \,.
\end{multline}
Using this expansion and the following integrals
\begin{equation} \label{integrale solido}
   \int \mathrm{d} o (1-\check{\bbm[k]}^2_i)= \frac{8}{3}\pi \, ,
   \qquad \mathrm{for} \;i=x,y,z \,,
\end{equation}
\begin{equation} \label{integrale solido2}
   \int \mathrm{d} o (1-\check{\bbm[k]}^2_i)X=0
      \qquad \mathrm{for} \;i=x,y,z \,,
\end{equation}
in Eq.\,(\ref{fase continuo2}), we obtain up to first order in
$v_0/c$
\begin{flalign}\label{fattore di fase}
  \xi(p,t)&=\frac{2 e^2 p^2}{3 \pi \hslash m_0^2 c^3 } \int_0^\infty \frac{\mathrm{d}\omega}{\omega}
  \exp \left(-\frac{\omega}{\Omega}\right)
  \left(\omega t-\sin\omega t\right)
  \nonumber \\ &=\frac{2\alpha p^2}{3\pi
  m_0^2 c^2
  }(\Omega t -\arctan \Omega t )\,,
\end{flalign}
where $\alpha=e^2/\hslash c$ is a dimensionless coupling constant.

\subsection{Decoherence function}
To obtain the explicit expression of $\Gamma^{p,p'}(t)$
(\ref{funzione di decoerenza calcolata2}) it is necessary to
calculate the trace on the field
\begin{equation}\label{funzione di Wigner}
  \chi (\gamma_{k,j}^{p,p'},\gamma_{k,j}^{p,p'}*)=\mathrm{tr}_F \left\{\exp
\left(\hat{\mathrm{a}}^{\dag}_{k,j}\gamma_{k,j}^{p,p'}-\hat{\mathrm{a}}_{k,j}\gamma_{k,j}^{p,p'}*\right)
  \hat{\rho}_F \right\} \,.
\end{equation}
The operator $\exp
\left(\hat{\mathrm{a}}^{\dag}_{k,j}\gamma_{k,j}^{p,p'}-\hat{\mathrm{a}}_{k,j}\gamma_{k,j}^{p,p'}*\right)$
is the generator of the coherent states of amplitude
$\gamma_{k,j}^{p,p'}$. It has been shown \cite{Privman-Mozyrsky
1998, Petruccione-Breuer libro 2002} that Eq.\,{\ref{funzione di
Wigner}} can be put in the form
\begin{equation}\label{funzione di Wigner2}
  \chi (\gamma_{k,j}^{p,p'},\gamma_{k,j}^{p,p'}*)=\exp \left\{-\frac{1}{2}|\gamma_{k,j}^{p,p'}|^2
  \mathrm{tr}_F \!\left[ \{\hat{\mathrm{a}}_{k,j},\hat{\mathrm{a}}^{\dag}_{k,j}\}\hat{\rho}_F
  \right]\right\} .
\end{equation}
Using Eqs.\,(\ref{funzione di Wigner}) and (\ref{funzione di
Wigner2}), the relation \cite{Petruccione-Breuer libro 2002}
\begin{equation}
  \mathrm{tr}_F \left[ \{\hat{\mathrm{a}}_{k,j},\hat{\mathrm{a}}^{\dag}_{k,j}\}\hat{\rho}_F
  \right]=\coth \left(\frac{\hslash \omega_k}{2k_B T}\right)
\end{equation}
obtained in the case of thermal distribution for $\hat{\rho}_F$
and
\begin{equation}
  |\gamma_{k,j}^{p,p'}|^2=\left[ (\bbm[p]-\bbm[p]\,')\cdot \varepsilon_{k,j} \right]^2
   \frac{\left[1-\cos ( \omega_k-\bbm[k]\cdot\bbm[v]_0)t\right]}{\hslash \omega_k( \omega_k-\bbm[k]\cdot\bbm[v]_0)^2}
  \frac{4\pi \, e^2 }{m_0^2V} \,,
\end{equation}
derived from the position following Eq.\,(\ref{funzione di
decoerenza calcolata2}) and using Eqs.\,(\ref{coefficiente di
accoppiamento}) and (\ref{alfa}), the decoherence function
(\ref{funzione di decoerenza calcolata2}) can be put in the form
\begin{multline}
  \Gamma^{p,p'}(t)=\frac{2\pi \, e^2 }{\hslash m_0^2
  V} \sum_k \frac{\left[1-\cos ( \omega_k-\bbm[k]\cdot\bbm[v]_0)t\right]}
  {\omega_k( \omega_k-\bbm[k]\cdot\bbm[v]_0)^2}\\  \times \coth
  \left(\frac{\hslash \omega_k}{2k_B T}\right) \sum_j
  \left[ (\bbm[p]-\bbm[p]\,')\cdot \varepsilon_{k,j} \right]^2\,.
\end{multline}
Taking the continuum limit on the field modes $\sum_k \rightarrow
V\int_0^\infty \mathrm{d}^3k /(2\pi)^3$, using $\sum_j \left[
(\bbm[p]-\bbm[p]\,')\cdot \varepsilon_{k,j}
\right]^2=(\bbm[p]-\bbm[p]\,')^2-\left[ (\bbm[p]-\bbm[p]\,')\cdot
\check{\bbm[k]} \right]^2$, inserting the cut off factor $\exp
(-\omega/\Omega)$ and introducing the variable $X$ defined in
Eq.\,(\ref{kappavizero}), we obtain
\begin{flalign}\label{parteraledecoerenzacontinuo}
  \Gamma^{p,p'}&(t)=\frac{e^2 }{4\pi^2 \hslash
  m_0^2c^3}\int_0^\infty \frac{\mathrm{d} \omega}{\omega} \exp \left(-\frac{\omega}{\Omega}\right) \coth
  \left(\frac{\hslash \omega}{2k_BT}\right) \nonumber \\ & \times
\int \mathrm{d}o \frac{1-\cos \left[ \omega
t(1-X)\right]}{(1-X)^2}
 \left[(\bbm[p]-\bbm[p]\,')_x^2(1-\check{\bbm[k]}^2_x) \right. \nonumber \\&  \left.+(\bbm[p]-\bbm[p]\,')_y^2(1-\check{\bbm[k]}^2_y)+
(\bbm[p]-\bbm[p]\,')_z^2(1-\check{\bbm[k]}^2_z)\right] \, ,
\end{flalign}
where, as before, $\mathrm{d}o$ is the infinitesimal solid angle.
\\
Indicating with $g(X)$ the $X$ dependent part within the integrand
in Eq.\,(\ref{parteraledecoerenzacontinuo}), for small values of
$v_0/c$, this can be expanded with respect to $X$ obtaining up to
the first order in $X$
\begin{equation}\label{espansione2}
  g(X)=\frac{1-\cos \left[ \omega
t(1-X)\right]}{(1-X)^2}\approx 1-\cos \omega t + g'(0)X \,.
\end{equation}
Using this expansion and Eqs.\,(\ref{integrale solido}) and
(\ref{integrale solido2}) to compute the angular integral in
Eq.\,(\ref{parteraledecoerenzacontinuo}), we obtain up to first
order in $v_0/c$
\begin{multline}\label{parteraledecoerenzacontinuofinale}
  \Gamma^{p,p'}(t)=\frac{2\alpha }{3\pi}\frac{(\bbm[p]-\bbm[p]\,')^2}{m_0^2c^2}
  \int_0^\infty \frac{\mathrm{d} \omega}{\omega} \exp
  \left(-\frac{\omega}{\Omega}\right)
  \\ \times (1-\cos \omega t)\coth
  \left(\hslash \omega/2k_BT\right) \,.
\end{multline}
Before  carrying out the frequency integral in
Eq.\,(\ref{parteraledecoerenzacontinuofinale}), we separate
$\Gamma^{p,p'}(t)$ in two parts,
$\Gamma^{p,p'}(t)=\Gamma^{p,p'}_{vac}(t)+\Gamma^{p,p'}_{th}(t)$: a
temperature independent part due to vacuum fluctuations and a
dependent one due to the thermal bath properties, which goes to
zero for $T \rightarrow 0$. From
Eq.\,(\ref{parteraledecoerenzacontinuofinale}) we obtain the
temperature independent contribution as
\begin{flalign}\label{contributo di vuoto0}
  \Gamma^{p,p'}_{vac}(t)&=\frac{2\alpha }{3\pi}\frac{(\bbm[p]-\bbm[p]\,')^2}{m_0^2c^2}
  \int_0^\infty \mathrm{d} \omega \exp \left(-\frac{\omega}{\Omega}\right)\frac{(1-\cos \omega t)
  }{\omega} \nonumber \\ &=\frac{2\alpha}{3\pi}\frac{(\bbm[p]-\bbm[p]\,')^2}{m_0^2c^2}\ln\sqrt{1 +
  \Omega^2t^2}\,,
  \end{flalign}
and the thermal contribution as
\begin{flalign}\label{contributo termico 0}
  \Gamma^{p,p'}_{th}(t)=& \frac{2\alpha }{3\pi}\frac{(\bbm[p]-\bbm[p]\,')^2}{m_0^2c^2}
  \int_0^\infty \frac{\mathrm{d} \omega}{\omega}
  \exp \left(-\frac{\omega}{\Omega}\right)
  (1-\cos \omega t) \nonumber \\ & \times [\coth  \left(\hslash \omega/2k_BT\right)-1] \nonumber
  \\
  =& \frac{2\alpha }{3\pi}\frac{(\bbm[p]-\bbm[p]\,')^2}{m_0^2c^2}
  \frac{1}{\beta}\int_0^t \mathrm{d} s \int_0^\infty
  \mathrm{d}x \exp \left(-\frac{k_BTx}{\hslash\Omega} \right)\nonumber \\ & \times [\coth  \left(x/2
  \right)-1]\sin(s x/\beta) \, .
\end{flalign}
For $k_BT\ll \hslash\Omega$ and introducing $\tau_F=\hslash/ \pi
k_BT
 \approx 2.43\cdot 10^{-12} s/T[K]$, we find
\begin{flalign}\label{contributo termico0,5}
  \Gamma^{p,p'}_{th}(t)
  \approx & \frac{2\alpha }{3\pi}\frac{(\bbm[p]-\bbm[p]\,')^2}{m_0^2c^2}
  \frac{1}{\beta}\int_0^t \mathrm{d} s \int_0^\infty
  \mathrm{d}x  \nonumber \\ & \times [\coth  \left(x/2
  \right)-1]\sin(s x/\beta) \nonumber \\
  =& \frac{2\alpha }{3\pi}\frac{(\bbm[p]-\bbm[p]\,')^2}{m_0^2c^2}
  \ln \left[ \frac{\sinh(t/\tau_F)}{t/\tau_F}   \right] \, ,
\end{flalign}
where in the integration on $x$ we have used the formula
\begin{equation}
  \int_0^\infty
  \mathrm{d}x [\coth  \left(x/2
  \right)-1]\sin(s x/\beta)=\pi\coth(\pi s/\beta) -
  \frac{\beta}{s}\,.
\end{equation}
Summing the vacuum contribution given by Eq.\,(\ref{contributo di
vuoto0}) and the thermal by Eq.\,(\ref{contributo termico0,5}), we
obtain for the decoherence function
\begin{flalign} \label{decoerenza complessiva0}
  \Gamma^{p,p'}(t)= \frac{2\alpha }{3\pi}\frac{(\bbm[p]-\bbm[p]\,')^2}{m_0^2c^2}
   \ln\left[\sqrt{1 + \Omega^2t^2} \, \;\frac{\sinh(t/\tau_F)}
  {t/\tau_F} \right]\,.
\end{flalign}

\section{} \label{appendice trasformata}

Here we report the explicit computation of the spatial reduced
density matrix that involves the double Fourier transform of the
reduced density matrix in the momentum space:
\begin{flalign}\label{C1}
 \rho^{r,r'}_S\!(t)\! =\!\frac{1}{(2\pi\hslash)^3} \!\!\int \!\! \mathrm{d}^3 p \,
  \mathrm{d}^3 p' \rho^{p,p'}_S\!(t) \exp \!\left[ \frac{i}{\hslash}(\bbm[p] \cdot \bbm[r] -
  \bbm[p]\,'\cdot \bbm[r] \, ')\right]\!.
\end{flalign}
Using Eq.\,(\ref{matricemomenti}) for $\rho^{p,p'}_S(t)$, with the
initial wave packet form of Eq.\,(\ref{roiniziale}), we can easily
decompose Eq.\,(\ref{C1}) in equal components:
\begin{flalign}
  &\rho^{r,r'}_S(t)  =\frac{\mathrm{N_x}I_x}{(2\pi\hslash)}\frac{\mathrm{N_y}I_y}{(2\pi\hslash)}
  \frac{\mathrm{N_z}I_z}{(2\pi\hslash)}\,,
\end{flalign}
where $x,y,z$ are mute indices, $\mathrm{N_x}=1/\sqrt{2\pi} \Delta
p_x $ and
\begin{flalign}
  &I_x =  \exp \left(-\frac{2 d^2 p_0^{x\,2}}{\hslash^2}\right) \int \mathrm{d} p_x \,
  \mathrm{d} p_x' \times \\ &
   \exp \left[-p_x^2\left(\Gamma-i \Phi+\frac{d^2}{\hslash^2}\right)
   +p_x\left(2\Gamma p_x'+\frac{2d^2p_0^x}{\hslash^2} +\frac{ix}{\hslash}\right) \right]  \nonumber \\ &
   \times \exp \left[-p_x'^2\left(\Gamma+i \Phi+\frac{d^2}{\hslash^2}\right)+p_x'\left(\frac{2d^2p_0^x}{\hslash^2}-\frac{ix'}{\hslash}\right)
  \right]\, , \nonumber
\end{flalign}
where we have introduced $\Delta r = \sqrt{3}d$, $\bbm[q]
=\bbm[r]-\bbm[r]_0$ of components  $(x,y,z)$ and we have
eliminated the explicit time dependence of $\Gamma$ and $\Phi$.
Using
\begin{equation}\label{integrale gaussiano campo2}
  \int_{-\infty}^{+\infty} \exp(-a x^2 + b x) \mathrm{d}
  x=\sqrt{\frac{\pi}{a}}\exp \left(\frac{b^2}{4 a} \right)  \:\:\: [\mathrm{Re}\;
  a > 0] \,,
\end{equation}
for the integral in $p_x$ we obtain
\begin{flalign}\label{I}
  &I_x = F(p_x') \frac{\sqrt{\pi} \exp \left[-
  p_0^{x\,2}d^2\frac{2\hslash^2(\Gamma-i \Phi +\frac{d^2}{\hslash^2})-d^2}
  {\hslash^4(\Gamma -i \Phi +\frac{d^2}{\hslash^2})}\right] }{\sqrt{\Gamma -i \Phi +\frac{d^2}{\hslash^2}}}
   \\
  &  \times\exp\left[-\frac{x^2}{4\hslash^2(\Gamma -i \Phi +\frac{d^2}{\hslash^2})}
 +i \frac{p_0^x d^2 x}{\hslash^3(\Gamma - i \Phi +\frac{d^2}{\hslash^2})}
\right]\,,
  \nonumber
\end{flalign}
where  $F(p_x')$ contains the integral in $p_x'$ and is equal to
\begin{flalign}
   &F(p_x')=\int_{-\infty}^{+\infty} \mathrm{d} p_x'
  \exp \left[-p_x'^2\frac{|\Gamma - i \Phi +\frac{d^2}{\hslash^2}|^2- \Gamma^2}{(\Gamma -i \Phi +\frac{d^2}{\hslash^2})}
  \right. \nonumber \\ & \left. \qquad\qquad\; +p_x'\left(\frac{2d^2p_0^x}{\hslash^2}-\frac{ix'}{\hslash}
  +\frac{i\hslash \Gamma x+2p_0^x d^2 \Gamma}{\hslash^2(\Gamma -i \Phi +\frac{d^2}{\hslash^2})}\right)
  \right]
  \nonumber \\
  &\qquad\;\:=
  \sqrt{\frac{\pi(\Gamma -i \Phi +\frac{d^2}{\hslash^2})}{2 \Gamma \frac{d^2}{\hslash^2}+\frac{d^4}{\hslash^4}+\Phi^2}}\times
   \\
  &  \exp \!
  \frac{\left[2p_0^xd^2(2\Gamma +\frac{d^2}{\hslash^2}-i \Phi) \!-i \hslash(\Gamma -i \Phi
+\frac{d^2}{\hslash^2})x'+i\hslash \Gamma x \right]^2}
  {4 \hslash^4(|\Gamma -i\Phi +\frac{d^2}{\hslash^2}|^2- \Gamma^2)(\Gamma -i \Phi +\frac{d^2}{\hslash^2})}. \nonumber
\end{flalign}
Substituting this result in Eq.\,(\ref{I}), simplifying and
rationalizing where it occurs, and posing the adimensional
quantity $Z= 1 + 2 \Gamma \hslash^2/ d^2+  \Phi^2 \hslash^4/d^4$,
we obtain after a lengthy calculation
\begin{flalign}
  &I_x = \!\frac{\pi \hslash ^2}{\sqrt{Z}d^2}
 \exp \! \left[-   \frac{p_0^{x\,2}d^2  (2\Gamma -i2 \Phi +\frac{d^2}{\hslash^2})(\Gamma +i \Phi +\frac{d^2}{\hslash^2})}
  {\hslash^2(\Gamma^2+Z\frac{d^4}{\hslash^4})}\right]
  \nonumber \\
  & \times \exp \left[\frac{p_0^{x\,2} (2\Gamma -i\Phi +\frac{d^2}{\hslash^2})^2(\Gamma +i\Phi +\frac{d^2}{\hslash^2})}
  {(\Gamma^2+Z\frac{d^4}{\hslash^4})Z} +\frac{2 \Gamma
  xx'}{4Zd^4/\hslash^2}\right]
 \nonumber \\
  &\times  \exp \left[-\frac{(\Gamma +i\Phi +\frac{d^2}{\hslash^2})x^2 +(\Gamma-i\Phi+\frac{d^2}{\hslash^2})x'^2 }
  {4Zd^4/\hslash^2}
  \right]
   \\
  & \times
  \exp \left[ i \frac{(2\Gamma +i\Phi +\frac{d^2}{\hslash^2}) p_0^x x -(2\Gamma-i\Phi+\frac{d^2}{\hslash^2}) p_0^x
  x'}  {Zd^2/\hslash}\right] \nonumber \,.
\end{flalign}
After some passage, from $I_x$ we obtain $\rho^{r,r'}_S(t)\! =
\mathrm{N}I_xI_yI_z/(2\pi\hslash)^3$ put in the form
\begin{flalign}\label{rospaziale2}
  &\rho^{r,r'}_S(t) =\frac{\mathrm{N}\hslash^3}{8d^6
  Z^{\frac{3}{2}}}\exp\left[ i \frac{   (1+2 \Gamma \hslash^2/d^2) \bbm[p]_0 \cdot
 (\bbm[q]-\bbm[q] \, ')}
  {\hslash Z} \right]  \nonumber  \\ & \times \exp \left\{- \frac{
  3[(\bbm[q]+2\bbm[p]_0 \Phi \hslash)^2+
  (\bbm[q] \, '+2\bbm[p]_0 \Phi \hslash )^2]}{4(3d^2Z)} \right\}
 \\
 & \times \exp \left\{ - \frac{3\Gamma \hslash^2(\bbm[q]-\bbm[q] \, ')^2/d^2 +i 3 \Phi \hslash^2(q^2-q'^2)/d^2}{4(3d^2Z)} \right\}
\nonumber \, .
\end{flalign}
The last can be put in a useful form to compute directly some
quantities, as
\begin{flalign}\label{rospaziale3}
   & \rho^{r,r'}_S(t) =\frac{\mathrm{N}\hslash^3}{8d^6 Z^{\frac{3}{2}}}\:
  \exp\left(-\frac{6p_0^2 \Phi^2 \hslash^2}
  {3d^2Z}\right)
  \\ &\times   \exp\left[i \frac{ (1+2 \Gamma \hslash^2/d^2)
  \bbm[p]_0 \cdot
  (\bbm[q]-\bbm[q] \, ') }
  {  \hslash Z}\right]  \nonumber
    \\ &
\times \exp\left[ - \frac{\frac{3}{2}(1+2 \Gamma
\hslash^2/d^2)(\bbm[q]-\bbm[q] \, ')^2+\frac{3}{2}(\bbm[q]+\bbm[q]
\, ')^2}{4(3d^2Z)} \right] \nonumber \\ & \times\exp\left[ \frac{
+12 \Phi \hslash \bbm[p]_0 \cdot (\bbm[q]+\bbm[q] \, ') -  i 3
\Phi \hslash^2(q^2-q'^2)/d^2 }
  {  4 (3 d^2Z)}\right]  \nonumber
\,.
\end{flalign}

\section{}\label{dinamica campo}

We compute the trace on the subsystem in the momentum basis,
$|p(t)\rangle$. In the interaction picture we have $|p(t)\rangle=
\exp [ itp^2/2m_0\hslash]|p(0)\rangle $, but because here we are
not interested in the free evolution we will use $|p(t)\rangle =
|p(0)\rangle$ for the trace, thus neglecting the phase factor
which isn't relevant for the following discussion.

We rewrite  the time evolution operator of Eq.\,(\ref{operatore
evoluzione cbh}) in the form
\begin{flalign} \label{operatore evoluzione conti sul campo}
  \hat{U}(t) \!=  \!\prod_{p,k,j} \exp \left[i  \xi(p,t)\hat{\sigma}_p\right] \exp \left[ \hat{\sigma}_p \left(
  \hat{\mathrm{a}}^{\dag}_{k,j}\beta_{k,j}^p-\hat{\mathrm{a}}_{k,j}\beta_{k,j}^{p*}\right)
  \right]  ,
\end{flalign}
where $\beta_{k,j}^p= g_{k,j}^p\alpha_k$ is given, using
Eqs.\,(\ref{coefficiente di accoppiamento}) and (\ref{alfa}),  by
\begin{equation}\label{beta}
  \beta_{k,j}^p= -\bbm[p]\cdot
\bbm[\varepsilon]_{k,j}\frac{e}{m_0}\sqrt{\frac{2\pi\hslash}{V
\omega_k }} \frac{1-\mathrm{e}^{i(\omega_k -
\bbm[k]\cdot\bbm[v]_0)t}}{\hslash (\omega_k -
\bbm[k]\cdot\bbm[v]_0)}\mathrm{e}^{-i \bbm[k]\cdot\bbm[r]_0} \,.
\end{equation}
Using Eq.\,(\ref{operatore evoluzione conti sul campo}) in
Eq.\,(\ref{elementi matrice ridotta campo}) we obtain
\begin{flalign}\label{elementi matrice ridotta campo1}
  & \hat{\rho}_F^{\lambda_{k,j},\lambda'_{k',j'}}(t)=\langle \lambda_{k,j}|\mathrm{tr}_S
  \left\{ \prod_{p_1,k_1,j_1}\prod_{p_2,k_2,j_2} \right. \nonumber \\ & \left\{ \exp \left[ \hat{\sigma}_{p_1} \left(
  \hat{\mathrm{a}}^{\dag}_{k_1,j_1}\beta_{k_1,j_1}^{p_1}-\hat{\mathrm{a}}_{k_1,j_1}\beta_{k_1,j_1}^{p_1*}
  \right) \right]
 \exp \left[i
\xi(p_1,t)\hat{\sigma}_{p_1}\right] \right.  \nonumber \\& \left.
\left. \times
  \sum_{p,p'}|p\rangle  |\{0_{\bar{k},\bar{j}}\}\rangle \langle \{0_{\dot{k},\dot{j}}\}| \langle p'|
  N_\varepsilon \delta^\varepsilon_{p,\bar{p}}
  \delta ^{\varepsilon '}_{p',\bar{p}}
  \right. \right.  \nonumber \\ & \times
   \exp \left[ \hat{\sigma}_{p_2} \left(-
  \hat{\mathrm{a}}^{\dag}_{k_2,j_2}\beta_{k_2,j_2}^{p_2}+\hat{\mathrm{a}}_{k_2,j_2}
  \beta_{k_2,j_2}^{p_2*}\right) \right]
  \nonumber \\ & \left.\left. \times
  \exp \left[-i
  \xi(p_2,t)\hat{\sigma}_{p_2}\right] \right\} \right\} |\lambda'_{k',j'}\rangle
  \,.
\end{flalign}
Using the cyclicity of the trace and
\begin{flalign}\label{generatore stato coerente}
  &\exp \hat{\sigma}_p \left(
  \hat{\mathrm{a}}^{\dag}_{k,j}\beta_{k,j}^p-\hat{\mathrm{a}}_{k,j}\beta_{k,j}^{p*}\right)
  |p\rangle|0_{k,j}\rangle \nonumber \\
  &=|p\rangle \exp  \left(
  \hat{\mathrm{a}}^{\dag}_{k,j}\beta_{k,j}^p-\hat{\mathrm{a}}_{k,j}\beta_{k,j}^{p*}\right)
  |0_{k,j}\rangle= |p\rangle |\beta_{k,j}^p\rangle \, ,
\end{flalign}
where the amplitude of the coherent state $|\beta_{k,j}^p\rangle$
depends on the momentum component $\bbm[p]$, Eq.\,(\ref{elementi
matrice ridotta campo1}) becomes
\begin{flalign}\label{elementi matrice ridotta campo2}
  & \hat{\rho}_F^{\lambda_{k,j},\lambda'_{k',j'}}(t)=\int \mathrm{d}^3 p'' \sum_{p,p'}\delta(p''-p) \delta(p''-p')
  \bar{N } \nonumber \\ & \times  \delta^\varepsilon_{p,\bar{p}}
  \delta ^{\varepsilon '}_{p',\bar{p}} \prod_{k_1,j_1}\prod_{k_2,j_2}  \langle \lambda_{k,j}
  |\beta_{k_1,j_1}^p \rangle
  \langle \beta_{k_2,j_2}^{p'}
  |\lambda'_{k',j'}\rangle \,.
\end{flalign}
Taking into account the explicit form of the scalar product
between coherent states
\begin{equation}
  \prod_{k',j'}  \langle \lambda_{k,j}
  |\beta_{k',j'}^p \rangle =\exp \left[-\frac{|\lambda_{k,j}|^2}{2}
  -\frac{|\beta_{k,j}^p|^2}{2}+\lambda_{k,j}^*\beta_{k,j}^p
  \right]\, ,
\end{equation}
Eq.\,(\ref{elementi matrice ridotta campo2}) becomes
\begin{multline}\label{elementi matrice ridotta campo3}
  \hat{\rho}_F^{\lambda_{k,j},\lambda'_{k',j'}}(t)=N_\varepsilon  \exp \left[-\frac{|\lambda_{k,j}|^2}{2}
  -\frac{|\lambda'_{k',j'}|^2}{2}-\frac{|\beta_{k,j}^{\bar{p}}|^2}{2}   \right. \\  \left.
  -\frac{|\beta_{k',j'}^{\bar{p}}|^2}{2}
  +\lambda^{*}_{k,j}\beta_{k,j}^{\bar{p}}
  +\lambda'_{k',j'}\beta_{k',j'}^{{\bar{p}}*} \right] \,,
\end{multline}
where the integral over the momenta has lead to the presence of
$\bar{\bbm[p]}$ in the $\beta_{k,j}^{\bar{p}}$.

Now, we calculate the average of the operator
$\hat{\mathrm{a}}^{\dag}_{k,j}\hat{\mathrm{a}}_{k,j}$ using the
trace in the coherent states basis, that for a generic operator
$\hat{A}$ has the form \cite{Privman-Mozyrsky 1998}
\begin{equation}\label{traccia su stati coerenti}
  \mathrm{Tr} \hat{A}\equiv \frac{1}{\pi}\int \mathrm{d}(\mathrm{Re}
  \lambda)\,\mathrm{d}(\mathrm{Im} \lambda)\langle \lambda|\hat{A}|
  \lambda\rangle\,.
\end{equation}
Using $\hat{\mathrm{a}}^{\dag}_{k,j}\hat{\mathrm{a}}_{k,j}$ as
operator in Eq.\,(\ref{traccia su stati coerenti}) we obtain
(omitting the pedici $k$ and $j$ in $\lambda$)
\begin{multline}\label{media fotoni1}
  \langle\hat{\mathrm{a}}^{\dag}_{k,j}\hat{\mathrm{a}}_{k,j}\rangle=\mathrm{Tr}_F
  \left(\hat{\mathrm{a}}^{\dag}_{k,j}\hat{\mathrm{a}}_{k,j}\hat{\rho}_F \right)=
  \mathrm{Tr}_F \left(\hat{\mathrm{a}}^{\dag}_{k,j}\hat{\rho}_F \hat{\mathrm{a}}_{k,j} \right) \\ -
  \mathrm{Tr}_F \left(\hat{\rho}_F \right)     =
  \frac{1}{\pi}\int \mathrm{d}(\mathrm{Re}
  \lambda)\,\mathrm{d}(\mathrm{Im}
  \lambda)\left(|\lambda|^2-1\right)\hat{\rho}_F^{\lambda,\lambda}\,,
\end{multline}
where we have used the commutation rules satisfied by the
operators $\hat{\mathrm{a}}^{\dag}_{k,j}$ and
$\hat{\mathrm{a}}_{k,j}$, and the action of these operators on the
coherent sates
\begin{eqnarray}
  \hat{\mathrm{a}}_{k,j}| \lambda\rangle =\lambda| \lambda\rangle
  \;\;\; \mathrm{and} \;\;\; \langle \lambda|\, \hat{\mathrm{a}}^{\dag}_{k,j}=\lambda^*\langle
  \lambda|\,.
\end{eqnarray}
Substituting the diagonal elements of Eq.\,(\ref{elementi matrice
ridotta campo3}) in Eq.\,(\ref{media fotoni1}), we obtain
\begin{flalign}\label{media fotonicalcolo}
  \langle\hat{\mathrm{a}}^{\dag}_{k,j}\hat{\mathrm{a}}_{k,j}\rangle=&
  \frac{1}{\pi}\int \mathrm{d}(\mathrm{Re}
  \lambda)\,\mathrm{d}(\mathrm{Im} \lambda)\left(|\lambda|^2-1\right)
      \\ & \times  \exp \left[ -|\lambda|^2-|\beta_{k,j}^{\bar{p}}|^2   +\lambda^*\beta_{k,j}^{\bar{p}}
  +\lambda\beta_{k,j}^{{\bar{p}}*} \right] \,. \nonumber
\end{flalign}
Posing $\mathrm{Re}\lambda= \lambda_1$ and
$\mathrm{Im}\lambda=\lambda_2$, Eq.\,(\ref{media fotonicalcolo})
can be put in the form
\begin{flalign}\label{media fotonicalcolo2}
  & \langle\hat{\mathrm{a}}^{\dag}_{k,j}\hat{\mathrm{a}}_{k,j}\rangle=
  \frac{\exp \left(-|\beta_{k,j}^{\bar{p}}|^2\right)}{\pi}\int \mathrm{d}\lambda_1\, \mathrm{d}\lambda_2
  \left(\lambda_1^2+\lambda_2^2-1\right) \times  \\ &
  \exp \! \left[ -\lambda_1^2+2\lambda_1\left( \mathrm{Re}\beta_{k,j}^{\bar{p}}\right)^2  \right]
  \exp \! \left[ -\lambda_2^2+2\lambda_2\left( \mathrm{Im}\beta_{k,j}^{\bar{p}}\right)^2
  \right]\nonumber .
\end{flalign}
The integrals involved in Eq.\,(\ref{media fotonicalcolo2}) are of
the Gaussian type (\ref{integrale gaussiano campo2}) and
\begin{equation}\label{integrale gaussiano campo}
  \int_{-\infty}^{+\infty} \! \mathrm{d}x\, x^2 \exp \!
  \left(-ax^2+bx\right)=\frac{\sqrt{\pi}(2a+b^2)}{4a^{5/2}}\exp \!
  \left(\frac{b^2}{4a^2}\right) .
\end{equation}
Using Eqs.\,(\ref{integrale gaussiano campo}) and (\ref{integrale
gaussiano campo2}) in Eq.\,(\ref{media fotonicalcolo}) we obtain
easily
\begin{multline}\label{media fotoni2}
  \langle\hat{\mathrm{a}}^{\dag}_{k,j}\hat{\mathrm{a}}_{k,j}\rangle=|\beta_{k,j}^{\bar{p}}|^2=
 \frac{4\pi e^2  \left(\bar{\bbm[p]}\cdot \bbm[\varepsilon]_{k,j}\right)^2}{Vm_0^2\hslash \omega^3_k(1 -
  X)^2}\\
 \times \left\{1-\cos \left[\omega_k t (1-X)\right]\right\} \,,
\end{multline}
with $X$ defined in  Eq.\,(\ref{kappavizero}). Eq.\,(\ref{media
fotoni2}) represents the average number of photons, of the mode of
the field represented by $\{k,j\}$, that compose the cloud
associated to the momentum $\bar{\bbm[p]}$. To obtain the trend of
the total number of photons, then, we must sum over the
polarizations and over $\bbm[k]$. Using Eq.\,(\ref{p scalare
epsilon}) to perform the sum over $j$ and
Eq.\,({\ref{kappavizero}}) we obtain
\begin{multline}\label{media fotoni3}
  \sum_j\langle\hat{\mathrm{a}}^{\dag}_{k,j}\hat{\mathrm{a}}_{k,j}\rangle=
  \left[{\bar{p}}^2-\left(\bar{\bbm[p]}\cdot \check{\bbm[k]}\right)^2  \right]
  \frac{4 \pi e^2}{Vm_0^2\hslash \omega_k^3}
  \\ \times \frac{1-\cos \left[\omega t (1-X)\right]}{(1-X)^2}\,.
\end{multline}
Performing directly the limit to continuum on the field modes and
inserting the usual cut off  factor $\exp(-\omega/\Omega)$, the
sum over the $\check{\bbm[k]}$ assumes the form
\begin{flalign}\label{media fotoni4}
  & \langle \hat{n}_{\bar{p}}\rangle =\frac{V}{(2\pi)^3}\int \mathrm{d}^3 k \exp\left(-\frac{\omega}{\Omega}\right)
  \sum_j\langle\hat{\mathrm{a}}^{\dag}_{k,j}\hat{\mathrm{a}}_{k,j}\rangle \nonumber  \\ &  =
  \frac{ e^2}{2\pi^2 m_0^2\hslash c^3}\int_0^\infty \frac{\mathrm{d}\omega}{\omega}\exp\left(-\frac{\omega}{\Omega}\right)
  \int \mathrm{d}\hat{\sigma} \frac{1-\cos \left[\omega t
  (1-X)\right]}{(1-X)^2}\nonumber
  \\  & \times
   \left[{\bar{p}}_x^2\left(1-\check{\bbm[k]}_x^2\right)+{\bar{p}}_y^2
  \left(1-\check{\bbm[k]}_y^2\right)+{\bar{p}}_z^2\left(1-\check{\bbm[k]}_z^2\right)
  \right] \,.
\end{flalign}
Using the expansion of Eq.\,(\ref{espansione2}), and
Eqs.\,(\ref{integrale solido}) and (\ref{integrale solido2}) in
Eq.\,(\ref{media fotoni4}), we obtain that the angular integral is
equal to $(1-\cos \omega t)8\pi \bar{p}^2/3$, while the resulting
integral over frequencies gives $\ln
\left(1+\Omega^2t^2\right)/2$. The average number of photons at
time $t$ that compose the cloud associated to momentum
$\bar{\bbm[p]}$ is then equal to
\begin{flalign}\label{media fotoni5}
  \langle\hat{n}_{\bar{p}}\rangle=
  \frac{2 \alpha }{3\pi }\frac{{\bar{p}}^2}{m_0^2 c^2}\ln
  \left(1+\Omega^2t^2\right) \,.
\end{flalign}

The average energy associated to this cloud of photons is obtained
multiplying the integrand of the final integral over the
frequencies (\ref{media fotoni4}) for $\hslash \omega$ and using
considerations similar to those leading from Eq.\,(\ref{media
fotoni4}) to (\ref{media fotoni5}):
\begin{flalign}\label{energia media fotoni}
  \langle \hat{E}_{F}\rangle& =\frac{V}{(2\pi)^3} \int \mathrm{d}^3 k
  \exp\left(-\frac{\omega}{\Omega}\right)
  \hslash\omega \sum_j\langle\hat{\mathrm{a}}^{\dag}_{k,j}\hat{\mathrm{a}}_{k,j}\rangle  \nonumber \\ &=
  \frac{4\alpha}{3\pi}\frac{\hslash {\bar{p}}^2}{m_0^2 c^2}\int_0^\infty
  \mathrm{d}\omega \exp \left(-\frac{\omega}{\Omega}\right)
  (1-\cos \omega t)\nonumber \\ & =
  \frac{8\alpha}{3\pi}\frac{\hslash \Omega }{m_0 c^2}
  \frac{\Omega^2t^2}{1+\Omega^2t^2}\frac{{\bar{p}}^2}{2m_0} \,.
\end{flalign}


\end{document}